\documentclass[copyright,creativecommons,sharealike]{eptcs} 
\providecommand{\event}{QAPL 2013}

\newif\ifdescdefetc

\newif\ifqaplsubmission

\newif\iftoeptcs
\toeptcstrue

\newif\ifnoappendix
\usepackage{currfile}
\ifcurrfilename{trpuc-paper.tex}{
\noappendixfalse
\qaplsubmissionfalse
\descdefetctrue
}{
\noappendixtrue
\qaplsubmissiontrue
\descdefetcfalse
}

\newif\ifforprint
\forprintfalse

\ifqaplsubmission

\else

\fi

\ifqaplsubmission
\else
\usepackage[style=alphabetic,backref,hyperref,maxbibnames=9999,mincrossrefs=2]{biblatex} 
\fi
\usepackage[ruled,linesnumbered,vlined]{algorithm2e}
\usepackage{amsmath}
\usepackage{amsthm}
\usepackage{color}
\usepackage{complexity}
\usepackage{dsfont}
\usepackage{latexsym}
\usepackage{paralist}
\usepackage{pgf}
\usepackage{placeins}
\usepackage{pifont}
\usepackage[figuresright]{rotating}
\usepackage{tikz}
\usetikzlibrary{arrows}
\usepackage{xspace}

\definecolor{darkgreen}{rgb}{0.0,0.33,0.0}

\def\ourtitle{%
Enhancing Unsatisfiable Cores for LTL with Information on Temporal Relevance%
}
\def\ourauthor{%
Viktor Schuppan%
}
\ifqaplsubmission
\def\oursubject{%
Eleventh International Workshop on Quantitative Aspects of Programming Languages and Systems, QAPL 2013, Rome, Italy, March 23-24, 2013%
}
\else
\def\oursubject{%
Eleventh International Workshop on Quantitative Aspects of Programming Languages and Systems, QAPL 2013, Rome, Italy, March 23-24, 2013 (full version)%
}
\fi
\def\ourkeywords{%
LTL, unsatisfiable cores, vacuity, temporal resolution, Parikh image%
}
\ifforprint
\hypersetup{
  pdfauthor = {\ourauthor},
  pdftitle = {\ourtitle},
  pdfsubject = {\oursubject},
  pdfkeywords = {\ourkeywords},
  pdfcreator = {LaTeX with hyperref package},
  colorlinks = true,
  linkcolor = black,
  citecolor = black,
  filecolor = black,
  urlcolor = black,
  pdfproducer = {pdflatex},
  bookmarksnumbered = true
}
\else
\hypersetup{
  pdfauthor = {\ourauthor},
  pdftitle = {\ourtitle},
  pdfsubject = {\oursubject},
  pdfkeywords = {\ourkeywords},
  pdfcreator = {LaTeX with hyperref package},
  colorlinks = true,
  citecolor = darkgreen,
  urlcolor = blue,
  pdfproducer = {pdflatex},
  bookmarksnumbered = true
}
\fi

\ifqaplsubmission
\else
\makeatletter
\defbibheading{bibintoc}[\refname]{%
\section*{#1}%
\addcontentsline{toc}{section}{#1}%
\markboth{\hfill\titlerunning}{\authorrunning\hfill}}%
\makeatother
\fi

\ifqaplsubmission
\else
\DeclareBibliographyDriver{incollection}{%
  \usebibmacro{bibindex}%
  \usebibmacro{begentry}%
  \usebibmacro{author/translator+others}%
  \setunit{\labelnamepunct}\newblock
  \usebibmacro{title}%
  \newunit
  \printlist{language}%
  \newunit\newblock
  \usebibmacro{byauthor}%
  \newunit\newblock
  \usebibmacro{in:}%
  \iffieldundef{crossref}
    {\usebibmacro{incollection:crossref:full}}
    {\usebibmacro{crossref:label}}
  \newunit\newblock
  \usebibmacro{chapter+pages}%
  \iffieldundef{crossref}
    {\usebibmacro{crossref:extrainfo1}}
    {}
  \newunit\newblock
  \usebibmacro{doi+eprint+url}%
  \iffieldundef{crossref}
    {\usebibmacro{crossref:extrainfo2}}
    {}
  \setunit{\bibpagerefpunct}\newblock
  \usebibmacro{pageref}%
  \usebibmacro{finentry}}

\newbibmacro{incollection:crossref:full}{%
  \usebibmacro{maintitle+booktitle}%
  \newunit\newblock
  \usebibmacro{byeditor+others}%
  \newunit\newblock
  \printfield{edition}%
  \newunit
  \iffieldundef{maintitle}
    {\printfield{volume}%
     \printfield{part}}
    {}%
  \newunit
  \printfield{volumes}%
  \newunit\newblock
  \usebibmacro{series+number}%
  \newunit\newblock
  \printfield{note}%
  \newunit\newblock
  \printlist{organization}%
  \newunit
  \usebibmacro{publisher+location+date}}

\DeclareBibliographyDriver{inproceedings}{%
  \usebibmacro{bibindex}%
  \usebibmacro{begentry}%
  \usebibmacro{author/translator+others}%
  \setunit{\labelnamepunct}\newblock
  \usebibmacro{title}%
  \newunit
  \printlist{language}%
  \newunit\newblock
  \usebibmacro{byauthor}%
  \newunit\newblock
  \usebibmacro{in:}%
  \iffieldundef{crossref}
    {\usebibmacro{inproceedings:crossref:full}}
    {\usebibmacro{crossref:label}}
  \newunit\newblock
  \usebibmacro{chapter+pages}%
  \iffieldundef{crossref}
    {\usebibmacro{crossref:extrainfo1}}
    {}
  \newunit\newblock
  \usebibmacro{doi+eprint+url}%
  \iffieldundef{crossref}
    {\usebibmacro{crossref:extrainfo2}}
    {}
  \setunit{\bibpagerefpunct}\newblock
  \usebibmacro{pageref}%
  \usebibmacro{finentry}}

\newbibmacro{inproceedings:crossref:full}{%
  \usebibmacro{maintitle+booktitle}%
  \newunit\newblock
  \usebibmacro{event+venue+date}%
  \newunit\newblock
  \usebibmacro{byeditor+others}%
  \newunit\newblock
  \iffieldundef{maintitle}
    {\printfield{volume}%
     \printfield{part}}
    {}%
  \newunit
  \printfield{volumes}%
  \newunit\newblock
  \usebibmacro{series+number}%
  \newunit\newblock
  \printfield{note}%
  \newunit\newblock
  \printlist{organization}%
  \newunit
  \usebibmacro{publisher+location+date}}

\newbibmacro{crossref:label}{%
  \entrydata{\strfield{crossref}}
     {\printtext{\mkbibbrackets
        {\printfield{labelalpha}\printfield{extraalpha}}}}}

\newbibmacro{crossref:extrainfo1}{%
  \newunit\newblock
  \iftoggle{bbx:isbn}
    {\printfield{isbn}}
    {}}%

\newbibmacro{crossref:extrainfo2}{%
  \newunit\newblock
  \usebibmacro{addendum+pubstate}}%
\fi

\ifqaplsubmission
\else
\ifforprint
\bibliography{trpuc-paper}
\else
\bibliography{trpuc-paper-url}
\fi
\fi

\ifqaplsubmission
\else

\fi

\newtheorem{definition}{Definition}
\newtheorem{theorem}{Theorem}
\newtheorem{proposition}{Proposition}
\newtheorem{lemma}{Lemma}
\newtheorem{remark}{Remark}

\newcommand{\mydefinition}[1]{\ifdescdefetc\begin{definition}[{#1}]\else\begin{definition}\fi}

\newcommand{\mytheorem}[4]{%
\newcommand{#1}[1]{%
\ifthenelse{\equal{##1}{true}}{\newcounter{cnt:#2}\setcounter{cnt:#2}{\value{theorem}}}{\newcounter{cnt:#2:backup}\setcounter{cnt:#2:backup}{\value{theorem}}\setcounter{theorem}{\value{cnt:#2}}}%
\ifdescdefetc\begin{theorem}[#3]\else\begin{theorem}\fi%
\ifthenelse{\equal{##1}{true}}{\label{#2}}{}%
{#4}%
\end{theorem}%
\ifthenelse{\equal{##1}{true}}{}{\setcounter{theorem}{\value{cnt:#2:backup}}}%
}%
}

\newcommand{\myproposition}[4]{%
\newcommand{#1}[1]{%
\ifthenelse{\equal{##1}{true}}{\newcounter{cnt:#2}\setcounter{cnt:#2}{\value{proposition}}}{\newcounter{cnt:#2:backup}\setcounter{cnt:#2:backup}{\value{proposition}}\setcounter{proposition}{\value{cnt:#2}}}%
\ifdescdefetc\begin{proposition}[#3]\else\begin{proposition}\fi%
\ifthenelse{\equal{##1}{true}}{\label{#2}}{}%
{#4}%
\end{proposition}%
\ifthenelse{\equal{##1}{true}}{}{\setcounter{proposition}{\value{cnt:#2:backup}}}%
}%
}

\newcommand{\mylemma}[4]{%
\newcommand{#1}[1]{%
\ifthenelse{\equal{##1}{true}}{\newcounter{cnt:#2}\setcounter{cnt:#2}{\value{lemma}}}{\newcounter{cnt:#2:backup}\setcounter{cnt:#2:backup}{\value{lemma}}\setcounter{lemma}{\value{cnt:#2}}}%
\ifdescdefetc\begin{lemma}[#3]\else\begin{lemma}\fi%
\ifthenelse{\equal{##1}{true}}{\label{#2}}{}%
{#4}%
\end{lemma}%
\ifthenelse{\equal{##1}{true}}{}{\setcounter{lemma}{\value{cnt:#2:backup}}}%
}%
}

\newcommand{\myremark}[4]{%
\newcommand{#1}[1]{%
\ifthenelse{\equal{##1}{true}}{\newcounter{cnt:#2}\setcounter{cnt:#2}{\value{remark}}}{\newcounter{cnt:#2:backup}\setcounter{cnt:#2:backup}{\value{remark}}\setcounter{remark}{\value{cnt:#2}}}%
\ifdescdefetc\begin{remark}[#3]\else\begin{remark}\fi%
\ifthenelse{\equal{##1}{true}}{\label{#2}}{}%
{#4}%
\end{remark}%
\ifthenelse{\equal{##1}{true}}{}{\setcounter{remark}{\value{cnt:#2:backup}}}%
}%
}

\newcommand{\vstodo}[1]%
{{\color{red} \bf \tt<VS todo>}{#1}{\color{red} \bf \tt</VS todo>}}

\newcommand{\uc}{UC\xspace}
\newcommand{\ucs}{UCs\xspace}
\newcommand{\UC}{UC\xspace}
\newcommand{\UCs}{UCs\xspace}

\newcommand{\tr}{TR\xspace}
\newcommand{\TR}{TR\xspace}

\newcommand{\result}[1]{{\emph{#1}}\xspace}
\newcommand{\sat}{\result{sat}}

\newcommand{\unsat}{\result{unsat}}
\newcommand{\Unsat}{\result{Unsat}}

\newcommand{\proofbiimplies}{\ensuremath{\Leftrightarrow}}
\newcommand{\definedas}{\ensuremath{\equiv}}

\newcommand{\mkpair}[2]{\ensuremath{({#1},{#2})}}
\newcommand{\mkpowerset}[1]{\ensuremath{2^{#1}}}

\newcommand{\fcardinality}[1]{\ensuremath{|{#1}|}}
\newcommand{\fmin}[1]{\ensuremath{min({#1})}}

\newcommand{\bigo}{\ensuremath{\mathcal{O}}}
\newcommand{\fbigo}[1]{\ensuremath{\bigo({#1})}}

\newcommand{\allaps}{\ensuremath{\mathit{AP}}}
\newcommand{\ap}{\ensuremath{\mathit{p}}}
\newcommand{\app}{\ensuremath{\mathit{q}}}
\newcommand{\appp}{\ensuremath{\mathit{r}}}
\newcommand{\apppp}{\ensuremath{\mathit{s}}}
\newcommand{\apres}{\ensuremath{\mathit{l}}}

\newcommand{\dap}{\ensuremath{\mathit{P}}}
\newcommand{\dapp}{\ensuremath{\mathit{Q}}}
\newcommand{\dappp}{\ensuremath{\mathit{R}}}
\newcommand{\dapppp}{\ensuremath{\mathit{S}}}

\newcommand{\allbools}{\ensuremath{\mathds{B}}}
\newcommand{\false}{\ensuremath{0}}
\newcommand{\true}{\ensuremath{1}}

\newcommand{\iword}{\ensuremath{\pi}}
\newcommand{\fiwordgetpos}[2]{\ensuremath{{#1}[{#2}]}}

\newcommand{\allnats}{\ensuremath{\mathds{N}}}
\newcommand{\setpos}{\ensuremath{I}}
\newcommand{\setposp}{\ensuremath{I'}}
\newcommand{\setpospp}{\ensuremath{I''}}

\newcommand{\pos}{\ensuremath{{i}}}
\newcommand{\posp}{\ensuremath{{i'}}}
\newcommand{\pospp}{\ensuremath{{i''}}}

\newcommand{\maxind}{\ensuremath{{n}}}
\newcommand{\maxindp}{\ensuremath{{n'}}}
\newcommand{\maxindpp}{\ensuremath{{n''}}}
\newcommand{\maxindppp}{\ensuremath{{n'''}}}
\newcommand{\nat}{\ensuremath{{m}}}
\newcommand{\natp}{\ensuremath{{m'}}}

\newcommand{\inp}{\ensuremath{\mathit{\phi}}}
\newcommand{\inpp}{\ensuremath{\mathit{\phi'}}}
\newcommand{\inppp}{\ensuremath{\mathit{\phi''}}}

\newcommand{\inpuc}{\ensuremath{\mathit{\phi^{uc}}}}

\newcommand{\prt}{\ensuremath{\mathit{\psi}}}
\newcommand{\prtp}{\ensuremath{\mathit{\psi'}}}

\newcommand{\positivepolarity}{\ensuremath{+}}
\newcommand{\negativepolarity}{\ensuremath{-}}

\newcommand{\fbnotname}{\ensuremath{\neg}}
\newcommand{\fbnot}[1]{\ensuremath{\fbnotname{#1}}}
\newcommand{\fborname}{\ensuremath{\vee}}
\newcommand{\fbor}[2]{\ensuremath{{#1}\fborname{#2}}}
\newcommand{\fbandname}{\ensuremath{\wedge}}
\newcommand{\fband}[2]{\ensuremath{{#1}\fbandname{#2}}}
\newcommand{\fbimpliesname}{\ensuremath{\rightarrow}}
\newcommand{\fbimplies}[2]{\ensuremath{{#1}\fbimpliesname{#2}}}

\newcommand{\fbbiimpliesname}{\ensuremath{\leftrightarrow}}
\newcommand{\fbbiimplies}[2]{\ensuremath{{#1}\fbbiimpliesname{#2}}}

\newcommand{\fnextname}{\ensuremath{{\bf X}}}
\newcommand{\fnext}[1]{\ensuremath{\fnextname{#1}}}
\newcommand{\ffinallyname}{\ensuremath{{\bf F}}}
\newcommand{\ffinally}[1]{\ensuremath{\ffinallyname{#1}}}
\newcommand{\fgloballyname}{\ensuremath{{\bf G}}}
\newcommand{\fglobally}[1]{\ensuremath{\fgloballyname{#1}}}
\newcommand{\funtilname}{\ensuremath{{\bf U}}}
\newcommand{\funtil}[2]{\ensuremath{{#1}{\funtilname}{#2}}}
\newcommand{\freleasesname}{\ensuremath{{\bf R}}}
\newcommand{\freleases}[2]{\ensuremath{{#1}{\freleasesname}{#2}}}
\newcommand{\fweakuntilname}{\ensuremath{{\bf W}}}
\newcommand{\fweakuntil}[2]{\ensuremath{{#1}{\fweakuntilname}{#2}}}

\newcommand{\allclauses}{\ensuremath{\mathds{C}}}

\newcommand{\ficlause}[1]{\ensuremath{({#1})}}
\newcommand{\fgnxclause}[2]{\ensuremath{(\fglobally{(\fbor{({#1})}{(\fnext{({#2})})})})}}
\newcommand{\fgnclause}[1]{\ensuremath{(\fglobally{({#1})})}}
\newcommand{\feclause}[2]{\ensuremath{(\fglobally{(\fbor{({#1})}{(\ffinally{({#2})})})})}}
\newcommand{\emptyclause}{\ensuremath{\Box}}
\newcommand{\clause}{\ensuremath{{c}}}
\newcommand{\clausep}{\ensuremath{{c'}}}
\newcommand{\clausepp}{\ensuremath{{c''}}}

\newcommand{\setclauses}{\ensuremath{{C}}}
\newcommand{\setclausesp}{\ensuremath{{C'}}}
\newcommand{\setclausespp}{\ensuremath{{C''}}}
\newcommand{\setclausesuc}{\ensuremath{{C^{uc}}}}

\newcommand{\mainpartition}{\ensuremath{M}}
\newcommand{\mainpartitionp}{\ensuremath{M'}}
\newcommand{\looppartition}{\ensuremath{L}}
\newcommand{\looppartitionp}{\ensuremath{L'}}

\newcommand{\mainorlooppartition}{\ensuremath{M\!L}}

\newcommand{\refinferencerule}[1]{{\scriptsize\fbox{{#1}}}\xspace}
\newcommand{\refinitii}{\refinferencerule{init-ii}}
\newcommand{\refinitin}{\refinferencerule{init-in}}
\newcommand{\refstepnn}{\refinferencerule{step-nn}}
\newcommand{\refstepnx}{\refinferencerule{step-nx}}
\newcommand{\refstepxx}{\refinferencerule{step-xx}}
\newcommand{\refloopitinitx}{\refinferencerule{BFS-loop-it-init-x}}
\newcommand{\refloopitinitn}{\refinferencerule{BFS-loop-it-init-n}}
\newcommand{\refloopitinitc}{\refinferencerule{BFS-loop-it-init-c}}
\newcommand{\refloopitsub}{\refinferencerule{BFS-loop-it-sub}}
\newcommand{\refloopconclusionone}{\refinferencerule{BFS-loop-conclusion1}}
\newcommand{\refloopconclusiontwo}{\refinferencerule{BFS-loop-conclusion2}}
\newcommand{\refaugone}{\refinferencerule{aug1}}
\newcommand{\refaugtwo}{\refinferencerule{aug2}}

\newcommand{\fapwaitfor}[1]{\ensuremath{w{#1}}}

\newcommand{\tool}[1]{{\tt #1}\xspace}
\newcommand{\lwb}{\tool{LWB}}
\newcommand{\nusmv}{\tool{NuSMV}}
\newcommand{\pltl}{\tool{pltl}}
\newcommand{\trp}{\tool{TRP++}}
\newcommand{\tspass}{\tool{TSPASS}}
\newcommand{\benchmark}[1]{{\bf #1}}

\newcommand{\algassign}[2]{\ensuremath{{#1} \leftarrow {#2}}}

\newcommand{\fdisjunionname}{\ensuremath{\uplus}}
\newcommand{\fdisjunion}[2]{\ensuremath{{{#1}}\fdisjunionname{{#2}}}}

\newcommand{\graph}{\ensuremath{{G}}}
\newcommand{\graphp}{\ensuremath{{G'}}}
\newcommand{\setvertices}{\ensuremath{V}}
\newcommand{\setverticesp}{\ensuremath{V'}}

\newcommand{\vertex}{\ensuremath{v}}
\newcommand{\vertexp}{\ensuremath{v'}}

\newcommand{\setedges}{\ensuremath{E}}
\newcommand{\setedgesp}{\ensuremath{E'}}

\newcommand{\vertexlabelingname}{\ensuremath{L_V}}
\newcommand{\fvertexlabeling}[1]{\ensuremath{\vertexlabelingname({#1})}}
\newcommand{\edgelabelingname}{\ensuremath{L_{E'}}}
\newcommand{\vertexlabelingtwoname}{\ensuremath{L'_{V'}}}
\newcommand{\fvertexlabelingtwo}[1]{\ensuremath{\vertexlabelingtwoname({#1})}}

\newcommand{\partitioningv}{\ensuremath{\mathcal{Q}^\setvertices}}
\newcommand{\mainpartitionv}{\ensuremath{M^\setvertices}}
\newcommand{\looppartitionv}{\ensuremath{L^\setvertices}}
\newcommand{\edge}{\ensuremath{e}}

\newcommand{\ppath}{\ensuremath{\pi}}

\newcommand{\ltli}{LTL\emph{p}\xspace}
\newcommand{\qltl}{QLTL\xspace}

\newcommand{\fbnoti}[2]{\ensuremath{{\underset{\;#2\;}{\fbnotname}}{#1}}}
\newcommand{\fbori}[4]{\ensuremath{{#1}{\underset{{\;#3,#4\;}}{\fborname}}{#2}}}
\newcommand{\fbandi}[4]{\ensuremath{{#1}{\underset{{\;#3,#4\;}}{\fbandname}}{#2}}}
\newcommand{\fbimpliesi}[4]{\ensuremath{{#1}{\underset{{\;#3,#4\;}}{\fbimpliesname}}{#2}}}

\newcommand{\fnexti}[2]{\ensuremath{{\underset{\;#2\;}{\fnextname}}{#1}}}
\newcommand{\ffinallyi}[2]{\ensuremath{{\underset{\;#2\;}{\ffinallyname}}{#1}}}
\newcommand{\fgloballyi}[2]{\ensuremath{{\underset{\;#2\;}{\fgloballyname}}{#1}}}
\newcommand{\funtili}[4]{\ensuremath{{#1}{\underset{\;#3,#4\;}{\funtilname}}{#2}}}
\newcommand{\freleasesi}[4]{\ensuremath{{#1}{\underset{\;#3,#4\;}{\freleasesname}}{#2}}}

\newcommand{\inpi}{\ensuremath{\mathit{\theta}}}

\newcommand{\inpqi}{\ensuremath{\mathit{\chi}}}
\newcommand{\inpuci}{\ensuremath{\mathit{\theta^{uc}}}}
\newcommand{\inppuci}{\ensuremath{\mathit{{\theta'}^{uc}}}}

\newcommand{\prti}{\ensuremath{\mathit{\tau}}}
\newcommand{\prtpi}{\ensuremath{\mathit{\tau'}}}

\newcommand{\fls}[2]{\ensuremath{{#1} \!\cdot\! \allnats + {#2}}}
\newcommand{\fslsone}[2]{\ensuremath{\{\fls{{#1}}{{#2}}\}}}
\newcommand{\fslssone}[2]{\ensuremath{\fls{{#1}}{{#2}}}}
\newcommand{\fslssoneone}[1]{\ensuremath{{{#1}} \!\cdot\! \allnats}}
\newcommand{\fslssonetwo}[1]{\ensuremath{\{{#1}\}}}

\newcommand{\zeroset}{\ensuremath{\{0\}}}
\newcommand{\evenset}{\ensuremath{\fslssoneone{2}}}
\newcommand{\oddset}{\ensuremath{\fslssone{2}{1}}}

\newcommand{\period}{\ensuremath{p}}

\newcommand{\offset}{\ensuremath{o}}

\newcommand{\ficlausei}[1]{\ensuremath{({#1})}}

\newcommand{\feclausei}[4]{\ensuremath{(\fgloballyi{(\fbori{({#1})}{(\ffinallyi{({#2})}{{#4}})}{{#3}}{{#3}})}{{#3}})}}
\newcommand{\fclausei}[2]{\ensuremath{{\underset{\;{#2}\;}{{#1}}}}}

\newcommand{\yes}{{\color{green} \ding{52}}\xspace}
\newcommand{\no}{{\color{red} \ding{54}}\xspace}
\newcommand{\na}{---\xspace}

\newcommand{\fdCNF}[1]{\ensuremath{\mathit{SNF}({#1})}}
\newcommand{\fdCNFaux}[1]{\ensuremath{\mathit{SNF}_\mathit{aux}({#1})}}
\newcommand{\alldCNFvars}{\ensuremath{X}}
\newcommand{\fdCNFvar}[1]{\ensuremath{{x}_{#1}}}
\newcommand{\fdCNFvarm}[1]{\ensuremath{\color{blue}\setlength\fboxsep{1pt}\fbox{${x}_{#1}$}}}
\newcommand{\dCNFvar}{\ensuremath{x}}
\newcommand{\dCNFvarp}{\ensuremath{x'}}
\newcommand{\dCNFconj}{\ensuremath{c}}

\newcommand{\alphabet}{\ensuremath{\Sigma}}
\newcommand{\letter}{\ensuremath{\sigma}}
\newcommand{\formlang}{\ensuremath{L}}
\newcommand{\word}{\ensuremath{w}}
\newcommand{\setwords}{\ensuremath{W}}
\newcommand{\fparikhimagename}{\ensuremath{\Psi}}
\newcommand{\fparikhimage}[2]{\ensuremath{\fparikhimagename({#1},{#2})}}

\ifnoappendix
\iftoeptcs
\newcommand{\refformalcoreextractioni}{App.~B of \cite{fullversion}\xspace}
\newcommand{\refmoreexamples}{App.~C of \cite{fullversion}\xspace}
\newcommand{\refmoreplots}{App.~D of \cite{fullversion}\xspace}
\else
\newcommand{\refformalcoreextractioni}{App.~\ref{full-sec:formal-coreextractioni} of \cite{fullversion}\xspace}
\newcommand{\refmoreexamples}{App.~\ref{full-sec:moreexamples} of \cite{fullversion}\xspace}
\newcommand{\refmoreplots}{App.~\ref{full-sec:moreplots} of \cite{fullversion}\xspace}
\fi
\else
\newcommand{\refformalcoreextractioni}{App.~\ref{sec:formal-coreextractioni}\xspace}
\newcommand{\refmoreexamples}{App.~\ref{sec:moreexamples}\xspace}
\newcommand{\refmoreplots}{App.~\ref{sec:moreplots}\xspace}
\fi

\iftoeptcs
\else
\ifnoappendix
\fi
\fi

\begin{document}

\ifnoappendix
\title{\ourtitle}
\else
\title{\ourtitle\\(full version; r\input{SVNVERSION}\hspace{-0.25em}, \today)}
\fi

\def\titlerunning{\ourtitle}

\author{%
\ourauthor \email{Viktor.Schuppan@gmx.de}%
}

\def\authorrunning{V. Schuppan}

\maketitle

\begin{abstract}
LTL is frequently used to express specifications in many domains such as embedded systems or business processes.
Witnesses can help to understand why an LTL specification is satisfiable, and a number of approaches exist to make understanding a witness easier.
In the case of unsatisfiable specifications unsatisfiable cores (\ucs), i.e., parts of an unsatisfiable formula that are themselves unsatisfiable, are a well established means for debugging.
However, little work has been done to help understanding a \uc of an unsatisfiable LTL formula.
In this paper we suggest to enhance a \uc of an unsatisfiable LTL formula with additional information about the time points at which the subformulas of the \uc are relevant for unsatisfiability.
For example, in $\fband{(\fglobally{\ap})}{(\fnext{\fbnot{\ap}})}$ the first occurrence of $\ap$ is really only ``relevant'' for unsatisfiability at time point 1 (time starts at time point 0).
We present a method to extract such information from the resolution graph of a temporal resolution proof of unsatisfiability of an LTL formula.
We implement our method in \trp, and we experimentally evaluate it. Source code of our tool is available.

\end{abstract}

\section{Introduction}
\label{sec:introduction}

\ifqaplsubmission
\else
\paragraph{Motivation}
\fi

LTL \ifqaplsubmission\else(e.g., \cite{APnueli-FOCS-1977,EEmerson-HandbookOfTheoreticalComputerScience-1990})\xspace\fi and its relatives are important specification languages for reactive systems (e.g., \cite{CEisnerDFisman-2006}) and for business processes (e.g., \cite{MPesicWVanDerAalst-BPM-2006}).
Experience in verification (e.g., \ifqaplsubmission\cite{IBeerSBenDavidCEisnerYRodeh-FMSD-2001}\else\cite{IBeerSBenDavidCEisnerYRodeh-FMSD-2001,OKupferman-CONCUR-2006}\fi) and in synthesis (e.g., \cite{RBloemSGallerBJobstmannNPitermanAPnueliMWeiglhofer-COCV-2007}) has lead to specifications themselves becoming objects of analysis.
Typically, a specification is expected to be satisfiable. If it turns out to be unsatisfiable, finding a reason for unsatisfiability can help with the ensuing debugging.
Given the sizes of specifications of real world systems (e.g., \cite{AChiappiniACimattiLMacchiORebolloMRoveriASusiSTonettaBVittorini-ICSE-2010}) automated support for determining a reason for unsatisfiability of a specification is crucial.
\ifqaplsubmission
Frequently, such reason for unsatisfiability is taken to be a part of the unsatisfiable specification that is by itself unsatisfiable; this is called an unsatisfiable core (\uc) (e.g., \cite{VSchuppan-SCP-2012,RBakkerFDikkerFTempelmanPWognum-IJCAI-1993}).
\else
Frequently, such reason for unsatisfiability is taken to be a part of the unsatisfiable specification that is by itself unsatisfiable (e.g., \cite{VSchuppan-SCP-2012,RBakkerFDikkerFTempelmanPWognum-IJCAI-1993,JChinneckEDravnieks-ORSAJournalOnComputing-1991}); this is called an unsatisfiable core (\uc) (e.g., \cite{VSchuppan-SCP-2012,EGoldbergYNovikov-DATE-2003,LZhangSMalik-DATE-2003,HHoos-UBritishColumbiaTR-1999}).
\fi

Less simplistic ways to examine an LTL specification $\inp$ exist \cite{IPillSSempriniRCavadaMRoveriRBloemACimatti-DAC-2006}, and understanding their results also benefits from availability of \ucs.
First, one can ask whether a certain scenario $\inpp$, given as an LTL formula, is permitted by $\inp$. That is the case iff $\fband{\inp}{\inpp}$ is satisfiable.
Second, one can check whether $\inp$ ensures a certain LTL property $\inppp$. $\inppp$ holds in $\inp$ iff $\fband{\inp}{\fbnot{\inppp}}$ is unsatisfiable.
In the first case, if the scenario turns out not to be permitted by the specification, a \uc can help to understand which parts of the specification and the scenario are responsible for that.
In the second case a \uc can show which parts of the specification imply the property. Moreover, if there are parts of the property that are not part of the \uc, then those parts of the property could be strengthened without invalidating the property in the specification; i.e., the property is vacuously satisfied (e.g., \ifqaplsubmission\cite{IBeerSBenDavidCEisnerYRodeh-FMSD-2001}\else\cite{IBeerSBenDavidCEisnerYRodeh-FMSD-2001,OKupfermanMVardi-STTT-2003,RArmoniLFixAFlaisherOGrumbergNPitermanATiemeyerMVardi-CAV-2003,AGurfinkelMChechik-TACAS-2004,DFismanOKupfermanSSheinvaldFaragyMVardi-HVC-2008,OKupferman-CONCUR-2006}\fi).

Trying to help users to understand counterexamples in verification, which are essentially witnesses to a satisfiable formula, is a well established research topic (see, e.g., \cite{IBeerSBenDavidHChocklerAOrniRTrefler-CAV-2009} for some references).
In particular, it is common to add information to a counterexample on which parts of a counterexample are relevant at which points in time (e.g., \cite{KRaviFSomenzi-TACAS-2004,IBeerSBenDavidHChocklerAOrniRTrefler-CAV-2009}).
According to \cite{IBeerSBenDavidHChocklerAOrniRTrefler-CAV-2009} such explanations are an integral part of every counterexample trace in IBM's verification platform RuleBase PE.
Checks for vacuous specifications, which are closely related to \ucs \cite{VSchuppan-SCP-2012}, are an important feature of industrial hardware verification tools (see, e.g., \cite{IBeerSBenDavidCEisnerYRodeh-FMSD-2001,RArmoniLFixAFlaisherOGrumbergNPitermanATiemeyerMVardi-CAV-2003}).
In the academic world \ucs are an important part of design methods for embedded systems (e.g., \cite{IPillSSempriniRCavadaMRoveriRBloemACimatti-DAC-2006}) as well as for business processes (e.g., \cite{AAwadRGoreZHouJThomsonMWeidlich-InformationSystems-2012}).
Despite this relevance of \ucs efforts to provide additional information in the context of \ucs or vacuity have remained isolated (e.g., \cite{JSimmondsJDaviesAGurfinkelMChechik-STTT-2010}).

\ifqaplsubmission
\else
\paragraph{Example}
\fi

In this paper we suggest to enhance \ucs for LTL with information on the time points at which its subformulas are relevant for unsatisfiability. As illustration we discuss the example from the abstract in more detail, shown in \eqref{ex:abstract} again.
\ifqaplsubmission
\else
\begin{equation}\label{ex:abstract}
\fband{(\fglobally{\ap})}{(\fnext{\fbnot{\ap}})}
\end{equation}
\fi
When \eqref{ex:abstract} is evaluated on some word $\iword$ according to standard semantics of LTL (see Sec.~\ref{sec:preliminaries}), \eqref{ex:abstract} and both of its conjuncts, $\fglobally{\ap}$ and $\fnext{\fbnot{\ap}}$, are evaluated at time point 0 (time starts at time point 0), the operand of the $\fgloballyname$ operator, $\ap$, is evaluated at all time points in $\allnats$, and the operand of the $\fnextname$ operator, $\fbnot{\ap}$, as well as its operand, $\ap$, are evaluated at time point 1.
We can include this information into \eqref{ex:abstract} by writing the set of time points at which an operand is evaluated directly below the corresponding operator.
Note that in this scheme there is no place for the set of time points at which \eqref{ex:abstract} itself is evaluated; however, \eqref{ex:abstract} (as any LTL formula) will always be evaluated only at time point 0, so this need not be spelled out explicitly.
We then obtain \eqref{ex:abstract:ltlp}.
\ifqaplsubmission
\else
\begin{equation}\label{ex:abstract:ltlp}
\fbandi{(\fgloballyi{\ap}{\allnats})}{(\fnexti{\fbnoti{\ap}{\fslssonetwo{1}}}{\fslssonetwo{1}})}{\zeroset}{\zeroset}
\end{equation}
\fi
It is easy to see that \eqref{ex:abstract} evaluates to $\false$ on any word $\iword$, i.e., it is unsatisfiable.
The reason for this is that at time point 1 $\ap$ would have to be $\true$ in the first conjunct $\fglobally{\ap}$ and $\false$ in the second conjunct $\fnext{\fbnot{\ap}}$.
Notice that for this to happen the operand of $\fglobally{\ap}$, $\ap$, needs to be evaluated only at time point 1; it is immaterial at any other time point.
\ifqaplsubmission
We can include this information into \eqref{ex:abstract:ltlp} by replacing $\allnats$ below $\fgloballyname$ with $\fslssonetwo{1}$, obtaining \eqref{ex:abstract:c}.
\else
\eqref{ex:abstract:ltlp} can be enhanced with this information by replacing $\allnats$ below $\fgloballyname$ with $\fslssonetwo{1}$, obtaining \eqref{ex:abstract:c}.
\fi
\eqref{ex:abstract:c} can be seen as a \uc of \eqref{ex:abstract}.
\ifqaplsubmission

{\scriptsize
\hspace{-0.06\linewidth}
\begin{minipage}{0.22\linewidth}
\begin{equation}\label{ex:abstract}
\fband{(\underset{\phantom{\zeroset}}{\fglobally{\ap}})}{(\fnext{\fbnot{\ap}})}
\end{equation}
\end{minipage}
\hspace{0.06\linewidth}
\begin{minipage}{0.31\linewidth}
\begin{equation}\label{ex:abstract:ltlp}
\fbandi{(\fgloballyi{\ap}{\allnats})}{(\fnexti{\fbnoti{\ap}{\fslssonetwo{1}}}{\fslssonetwo{1}})}{\zeroset}{\zeroset}
\end{equation}
\end{minipage}
\hspace{0.06\linewidth}
\begin{minipage}{0.31\linewidth}
\begin{equation}\label{ex:abstract:c}
\fbandi{(\fgloballyi{\ap}{\fslssonetwo{1}})}{(\fnexti{\fbnoti{\ap}{\fslssonetwo{1}}}{\fslssonetwo{1}})}{\zeroset}{\zeroset}
\end{equation}
\end{minipage}
}
\else
\begin{equation}\label{ex:abstract:c}
\fbandi{(\fgloballyi{\ap}{\fslssonetwo{1}})}{(\fnexti{\fbnoti{\ap}{\fslssonetwo{1}}}{\fslssonetwo{1}})}{\zeroset}{\zeroset}
\end{equation}
\fi

\ifqaplsubmission
\else
\paragraph{Approach}

We obtain the desired information as follows.
We use a temporal resolution-based \cite{MFisher-IJCAI-1991,MFisherCDixonMPeim-ACMTrComputationalLogic-2001} solver to determine unsatisfiability of a given LTL formula.
During the execution of the solver we construct a resolution graph \cite{VSchuppan-NFM-2013-full-arXiv}.
In the resolution graph we distinguish between edges where a premise is time-shifted 1 step into the future with respect to the conclusion and others where this is not the case.
An example for an inference that involves such a time step is concluding $\fglobally{(\ap)}$ from premises $\fglobally{(\fbor{\ap}{\fnext{\fbnot{\app}}})}$ and $\fglobally{(\app)}$: the second premise $\fglobally{(\app)}$ is implicitly converted to $\fglobally{(\fnext{\app})}$ to perform the inference.
I.e., the resulting edge from the second premise $\fglobally{(\app)}$ to the conclusion $\fglobally{(\ap)}$ involves a time step of 1, while this is not the case for the edge from the first premise $\fglobally{(\fbor{\ap}{\fnext{\fbnot{\app}}})}$ to the conclusion.
To determine the sets of time points at which a clause $\clause$ in an unsatisfiable set of SNF clauses (temporal resolution works on a clausal normal form called SNF \cite{MFisher-IJCAI-1991,MFisherPNoel-UManchesterTR-1992,MFisherCDixonMPeim-ACMTrComputationalLogic-2001}) is relevant for unsatisfiability we fix time point 0 as the time point at which the contradiction $\emptyclause$, which concludes the proof, is happening.
Then we count the number of time-shifts that occur on paths from $\clause$ to $\emptyclause$ in the resolution graph: $\clause$ is relevant for unsatisfiability at time point $\pos$ iff there exists a path from $\clause$ to $\emptyclause$ in the resolution graph that traverses exactly $\pos$ edges involving a time step of 1.
Note that the resolution graph may contain loops.
Therefore, to actually count the number of time-shifts on all possible paths from some clause $\clause$ to the contradiction $\emptyclause$, we regard the reversed resolution graph as a nondeterministic finite automaton over the language $\{0, 1\}$ with initial state $\emptyclause$ and final state $\clause$.
The desired information is then simply the Parikh image (e.g., \cite{ASalomaa-1973}) of the letter 1 in the language given by the automaton.
Parikh's theorem \cite{RParikh-JACM-1966} implies that the resulting sets of time points are semilinear.
To compute the Parikh images we use algorithms by Gawrychowski \cite{PGawrychowski-CIAA-2011} or Sawa \cite{ZSawa-FundamentaInformaticae-2013}.
For $\fcardinality{\setclauses}$ input SNF clauses and a resolution graph with $\fcardinality{\setverticesp}$ vertices backward reachable from $\emptyclause$ the information can be computed in time $\fbigo{\fcardinality{\setverticesp}^3 + \fcardinality{\setverticesp}^2 \cdot \fcardinality{\setclauses}}$.
Our experimental evaluation shows that the resulting overhead compared to extraction of \ucs without sets of time points is acceptable in practice.
\fi

\ifqaplsubmission
\else
\paragraph{Contributions}
\fi

In this paper we make the following contributions.
\begin{inparaenum}[(i)]
\item We suggest to enhance \ucs for LTL with information on the time points at which the subformulas of a \uc are relevant for unsatisfiability, leading to a more fine-grained notion of \ucs for LTL than \cite{VSchuppan-SCP-2012}.
\item We propose a method to obtain that information from a temporal resolution proof of unsatisfiability of an LTL formula.
\item We implement our method in \trp, and we experimentally evaluate it. We are not aware of any other tool that performs extraction of \ucs for propositional LTL at that level of granularity. We make the source code of our solver available.
\end{inparaenum}
Conceptually, under the frequently legitimate assumption that a system description can be translated into an LTL formula, our results extend to vacuity for LTL.

\ifqaplsubmission
\else
\paragraph{(Not) Just Debugging}
\fi

Besides debugging as outlined above \ucs are also used for avoiding the exploration of parts of a search space that can be known not to contain a solution for reasons ``equivalent'' to the reasons for previous failures (e.g., \ifqaplsubmission\cite{EClarkeMTalupurHVeithDWang-SAT-2003}\else\cite{EClarkeMTalupurHVeithDWang-SAT-2003,ACimattiMRoveriVSchuppanSTonetta-CAV-2007}\fi). While our results might also benefit that application, we focus on debugging.

\ifqaplsubmission
\else
\paragraph{Temporal Resolution as a Basis}
\fi

Temporal resolution \ifqaplsubmission\cite{MFisherCDixonMPeim-ACMTrComputationalLogic-2001}\else\cite{MFisher-IJCAI-1991,MFisherCDixonMPeim-ACMTrComputationalLogic-2001}\fi\xspace lends itself as a basis for enhancing \ucs for LTL with information on temporal relevance for two reasons.
First, the temporal resolution-based solver \trp \cite{UHustadtBKonev-CADE-2003,UHustadtBKonev-CollegiumLogicum-2004,trppptool} proved to be competitive in a recent evaluation of solvers for LTL satisfiability, in particular on unsatisfiable instances
\ifqaplsubmission
\cite{VSchuppanLDarmawan-ATVA-2011}.
\else
(see pp.~51--55 of the full version of \cite{VSchuppanLDarmawan-ATVA-2011}).
\fi
Second, a temporal resolution proof naturally induces a resolution graph \cite{VSchuppan-NFM-2013-full-arXiv}, which provides a clean framework for extracting information from the proof.
\ifqaplsubmission
\else
Note, that while the BDD-based solver \nusmv \cite{ACimattiEClarkeEGiunchigliaFGiunchigliaMPistoreMRoveriRSebastianiATacchella-CAV-2002} also performed well on unsatisfiable instances in \cite{VSchuppanLDarmawan-ATVA-2011}, the BDD layer makes extraction of information from the proof more involved.
On the other hand, the tableau-based solvers \lwb \cite{AHeuerdingGJaegerSSchwendimannMSeyfried-TABLEAUX-1995} and \pltl \cite{pltltool} provide access to a proof of unsatisfiability comparable to temporal resolution, yet tended to perform not as good on unsatisfiable instances in \cite{VSchuppanLDarmawan-ATVA-2011}.
\fi

\ifqaplsubmission
\else
\paragraph{Notion of Relevance}
\fi

In this paper we use the following notion of relevance.
Assume an LTL formula $\inp$ and a temporal resolution proof of its unsatisfiability.
Remove those parts of the proof that did not contribute to proving unsatisfiability.
Consider what is left to be relevant for unsatisfiability; this includes not only which subformulas of $\inp$ are used but also the time points at which they are used.
Clearly, this notion of relevance may not lead to results of minimal or minimum relevance: the fact that some part of $\inp$ is used at some time point in a specific proof of the unsatisfiability of $\inp$ does not mean that all such proofs will use that part of $\inp$ at that time point.
For other notions of relevance see, e.g., \cite{KRaviFSomenzi-TACAS-2004,IBeerSBenDavidHChocklerAOrniRTrefler-CAV-2009}.
The reasons for using our notion of relevance are pragmatic.
First, this notion of relevance can be computed with little overhead from a proof of unsatisfiability, which is assumed to be carried out anyway.
Second, the result of our notion of relevance can serve as an already reduced input to some of the other notions\ifqaplsubmission\else\xspace(e.g., \cite{AAwadRGoreZHouJThomsonMWeidlich-InformationSystems-2012,LZhangSMalik-SAT-2003,JChinneckEDravnieks-ORSAJournalOnComputing-1991})\fi.

\ifqaplsubmission
\else
\paragraph{Related Work}
\fi

Simmonds et al.~\cite{JSimmondsJDaviesAGurfinkelMChechik-STTT-2010} use SAT-based bounded model checking \ifqaplsubmission\else(e.g., \cite{ABiereACimattiEClarkeYZhu-TACAS-1999,ABiere-HandbookOfSatisfiability-2009})\xspace\fi for vacuity detection. They indicate which time points are relevant for showing that a variable is non-vacuous. They only consider $k$-step vacuity, i.e., taking into account bounded model checking runs up to a bound $k$, and leave the problem of removing the bound $k$ open.
\cite{VSchuppan-SCP-2012} mentions indicating time points at which subformulas of a \uc in LTL are relevant for unsatisfiability. The idea is not formalized. It first appears in the context of a \uc extraction algorithm that is complete only for a strict subset of LTL. Later \cite{VSchuppan-SCP-2012} proposes example \eqref{ex:everysecond} and conjectures that sets of time points can be obtained from a tableau method and that these sets are semilinear.
Some work \cite{KRaviFSomenzi-TACAS-2004,IBeerSBenDavidHChocklerAOrniRTrefler-CAV-2009} determines the time points at which propositions in witnesses of satisfiable LTL formulas are relevant for satisfiability.
\ifqaplsubmission
\else
For more general related work on \uc extraction see \cite{VSchuppan-NFM-2013-full-arXiv,VSchuppan-SCP-2012}.
\fi

\ifqaplsubmission
\else
\paragraph{Structure of the Paper}
\fi

This paper builds on a fair amount of previous work: we use temporal resolution as implemented in \trp \cite{MFisherCDixonMPeim-ACMTrComputationalLogic-2001,UHustadtBKonev-CADE-2003,UHustadtBKonev-CollegiumLogicum-2004,trppptool} and its extension to extract \ucs \cite{VSchuppan-NFM-2013-full-arXiv}.
To make this paper self-contained we provide a concise description of both.
However, to allow sufficient room for the contributions of this paper we have to limit the amount of explanation for previous work.
Temporal resolution has been developed since the early 1990s \cite{MFisher-IJCAI-1991}, and we refer to \cite{MFisherCDixonMPeim-ACMTrComputationalLogic-2001} for a general overview, to \ifqaplsubmission\cite{CDixon-AnnalsOfMathematicsAndArtificialIntelligence-1998,CDixon-ICTL-1997,CDixon-PhDThesis-1995}\else\cite{CDixon-AnnalsOfMathematicsAndArtificialIntelligence-1998,CDixon-ICTL-1997,CDixon-CADE-1996,CDixon-PhDThesis-1995}\fi\xspace for details on loop search, and to \cite{UHustadtBKonev-CADE-2003,UHustadtBKonev-CollegiumLogicum-2004,trppptool} for the implementation in \trp.
In \cite{VSchuppan-NFM-2013-full-arXiv} we provide some intuition on temporal resolution with a slant towards BDD-based symbolic model checking\ifqaplsubmission\else\xspace(e.g., \cite{JBurchEClarkeKMcMillanDDillLHwang-IaC-1992,EClarkeOGrumbergDPeled-2001})\fi.
Finally, we refer to Sec.~\ref{sec:coreextraction} for an example of an execution of the temporal resolution algorithm and the corresponding extraction of \ucs.

Section \ref{sec:preliminaries} starts with preliminaries.
Temporal resolution and its clausal normal form SNF are introduced in Sec.~\ref{sec:tr}.
In Sec.~\ref{sec:coreextraction} we restate the construction of a resolution graph and its use to obtain a \uc from \cite{VSchuppan-NFM-2013-full-arXiv}.
This is extended to compute time points at which subformulas are relevant for unsatisfiability in Sec.~\ref{sec:ltli}, \ref{sec:coreextractioni}.
A number of examples are spread throughout the paper; in Sec.~\ref{sec:example} we provide an example close to a real world situation.
We discuss our implementation and experimental evaluation in Sec.~\ref{sec:experimentalevaluation}.
Section \ref{sec:conclusions} concludes.
\ifqaplsubmission
Due to space constraints some proofs are sketched or omitted. For the full version of this paper \cite{fullversion} including proofs and for our implementation, examples, and log files see \cite{paperwebpage}.
\else
Due to space constraints some proofs are sketched or omitted in the main part; these can be found in the appendices. For our implementation, examples, and log files see \cite{paperwebpage}.
\fi

\section{Preliminaries}
\label{sec:preliminaries}

Let $\allnats$ be the naturals, and let $\setpos \subseteq \allnats$ be a set of naturals. $\setpos$ is \emph{linear} iff there exist two naturals $\period$ (\emph{period}) and $\offset$ (\emph{offset}) such that $\setpos = \fls{\period}{\offset}$. $\setpos$ is \emph{semilinear} iff it is the union of finitely many linear sets.
\ifqaplsubmission
\else

\fi
Let $\alphabet$ be a finite alphabet, $\letter \in \alphabet$ a letter in $\alphabet$, $\formlang \subseteq \alphabet^*$ a language over $\alphabet$, and $\word \in \formlang$ a word in $\formlang$. Define a function from words and letters to naturals $\fparikhimagename : \alphabet^* \times \alphabet \rightarrow \allnats, (\word, \letter) \mapsto \nat$ where $\nat$ is the number of occurrences of $\letter$ in $\word$. $\fparikhimagename$ is called \emph{Parikh mapping} and $\fparikhimage{\word}{\letter}$ is called the \emph{Parikh image} of $\letter$ in $\word$. The Parikh image of a set of words $\setwords$ is defined in the natural way: $\fparikhimage{\setwords}{\letter} = \{\fparikhimage{\word}{\letter} \mid \word \in \setwords\}$. Parikh's theorem \cite{RParikh-JACM-1966} states that for every context-free language $\formlang$, for every letter $\letter$, the Parikh image $\fparikhimage{\formlang}{\letter}$ is semilinear.\ifqaplsubmission\else \xspace See also \cite{ASalomaa-1973}.\fi

We use a standard version of LTL, see, e.g., \cite{EEmerson-HandbookOfTheoreticalComputerScience-1990}. Let $\allbools$ be the set of Booleans, and let $\allaps$ be a finite set of atomic propositions. The set of \emph{LTL formulas} is constructed inductively as follows. The Boolean constants $\false$ (false), $\true$ (true) $\in \allbools$ and any atomic proposition $\ap \in \allaps$ are LTL formulas. If $\prt$, $\prtp$ are LTL formulas, so are $\fbnot{\prt}$ (not), $\fbor{\prt}{\prtp}$ (or), $\fband{\prt}{\prtp}$ (and), $\fnext{\prt}$ (next time), $\funtil{\prt}{\prtp}$ (until), $\freleases{\prt}{\prtp}$ (releases), $\ffinally{\prt}$ (finally), and $\fglobally{\prt}$ (globally). We use $\fbimplies{\prt}{\prtp}$ (implies) as an abbreviation for $\fbor{\fbnot{\prt}}{\prtp}$\ifqaplsubmission\else\xspace, $\fbbiimplies{\prt}{\prtp}$ (equivalent) for $\fband{(\fbimplies{\prt}{\prtp})}{(\fbimplies{\prtp}{\prt})}$, and $\fweakuntil{\prt}{\prt}$ (weak until) for $\fbor{(\funtil{\prt}{\prtp})}{\fglobally{\prt}}$\fi. For the semantics of LTL see Tab.~\ref{tab:ltlsemantics}. An occurrence of a subformula $\prt$ of an LTL formula $\inp$ has \emph{positive polarity} ($\positivepolarity$) if it appears under an even number of negations in $\inp$ and \emph{negative polarity} ($\negativepolarity$) otherwise.
\begin{table}
\centering
{\tiny
\begin{tabular}{||lcl|lcl||}
\hline
\hline
$\mkpair{\iword}{\pos} \models \true, \not\models \false$ & &
&
$\mkpair{\iword}{\pos} \models \ap$ & $\proofbiimplies$ & $\ap \in \fiwordgetpos{\iword}{\pos}$
\\
\hline
$\mkpair{\iword}{\pos} \models \fbnot{\prt}$ & $\proofbiimplies$ & $\mkpair{\iword}{\pos} \not \models \prt$
&
$\mkpair{\iword}{\pos} \models \fnext{\prt}$ & $\proofbiimplies$ & $\mkpair{\iword}{\pos + 1} \models \prt$
\\
\hline
$\mkpair{\iword}{\pos} \models \fbor{\prt}{\prtp}$ & $\proofbiimplies$ & $\mkpair{\iword}{\pos} \models \prt \mbox{ or } \mkpair{\iword}{\pos} \models \prtp$
&
$\mkpair{\iword}{\pos} \models \fband{\prt}{\prtp}$ & $\proofbiimplies$ & $\mkpair{\iword}{\pos} \models \prt \mbox{ and } \mkpair{\iword}{\pos} \models \prtp$
\\
\hline
$\mkpair{\iword}{\pos} \models \funtil{\prt}{\prtp}$ & $\proofbiimplies$ & $\exists \posp \ge \pos \;.\; (\fband{\mkpair{\iword}{\posp} \models \prtp}{\forall \pos \le \pospp < \posp \;.\; \mkpair{\iword}{\pospp} \models \prt})$
&
$\mkpair{\iword}{\pos} \models \freleases{\prt}{\prtp}$ & $\proofbiimplies$ & $\forall \posp \ge \pos \;.\; (\fbor{\mkpair{\iword}{\posp} \models \prtp}{\exists \pos \le \pospp < \posp \;.\; \mkpair{\iword}{\pospp} \models \prt})$
\\
\hline
$\mkpair{\iword}{\pos} \models \ffinally{\prt}$ & $\proofbiimplies$ & $\exists \posp \ge \pos \;.\; \mkpair{\iword}{\posp} \models \prt$
&
$\mkpair{\iword}{\pos} \models \fglobally{\prt}$ & $\proofbiimplies$ & $\forall \posp \ge \pos \;.\; \mkpair{\iword}{\posp} \models \prt$
\\
\hline
\hline
\end{tabular}
}
\caption{\label{tab:ltlsemantics} Semantics of LTL. $\iword$ is a word in $(\mkpowerset{\allaps})^\omega$, $\pos$ is a time point in $\allnats$. $\iword$ satisfies $\inp$ iff $\mkpair{\iword}{0} \models \inp$.}
\end{table}

\section{Temporal Resolution (\TR) in \trp}
\label{sec:tr}

\ifqaplsubmission
\else
\begin{sloppypar}
\fi
\TR works on formulas in a clausal normal form called separated normal form (SNF) \ifqaplsubmission\cite{MFisherCDixonMPeim-ACMTrComputationalLogic-2001}\else\cite{MFisher-IJCAI-1991,MFisherPNoel-UManchesterTR-1992,MFisherCDixonMPeim-ACMTrComputationalLogic-2001}\fi.
For any atomic proposition $\ap \in \allaps$ $\ap$ and $\fbnot{\ap}$ are \emph{literals}.
Let $\ap_1, \ldots, \ap_\maxind$, $\app_1, \ldots, \app_\maxindp$, $\apres$ with $0 \le \maxind, \maxindp$ be literals such that $\ap_1, \ldots, \ap_\maxind$ and $\app_1, \ldots, \app_\maxindp$ are pairwise different. Then
\begin{inparaenum}[(i)]
\item $\ficlause{\fbor{\ap_1}{\fbor{\ldots}{\ap_\maxind}}}$ is an \emph{initial clause};
\item $(\fgloballyname((\fbor{\ap_1}{\fbor{\ldots}{\ap_\maxind}}) \fborname (\fnext{(\fbor{\app_1}{\fbor{\ldots}{\app_\maxindp}})})))$ is a \emph{global clause}; and
\item $\feclause{\fbor{\ap_1}{\fbor{\ldots}{\ap_\maxind}}}{\apres}$ is an \emph{eventuality clause}.
\end{inparaenum}
$\apres$ is called an \emph{eventuality literal}.
As usual an empty disjunction (resp.~conjunction) stands for $\false$ (resp.~\true).
${\ficlause{}}$ or ${\fgnclause{}}$, denoted $\emptyclause$, stand for $\false$ or $\fglobally{(\false)}$ and are called \emph{empty clause}.
The set of all SNF clauses is denoted $\allclauses$.
Let $\clause_1, \ldots, \clause_\maxind$ with $0 \le \maxind$ be SNF clauses. Then $\bigwedge_{1 \le \pos \le \maxind} \clause_\pos$ is an LTL formula in \emph{SNF}.
Every LTL formula $\inp$ can be transformed into an equisatisfiable formula $\inpp$ in SNF \ifqaplsubmission\cite{MFisherCDixonMPeim-ACMTrComputationalLogic-2001}\else\cite{MFisher-IJCAI-1991,MFisherPNoel-UManchesterTR-1992,MFisherCDixonMPeim-ACMTrComputationalLogic-2001}\fi.
\ifqaplsubmission
\else
\end{sloppypar}
\fi

We now describe \tr \ifqaplsubmission\cite{MFisherCDixonMPeim-ACMTrComputationalLogic-2001}\else\cite{MFisher-IJCAI-1991,MFisherCDixonMPeim-ACMTrComputationalLogic-2001}\fi\xspace as implemented in \trp \cite{UHustadtBKonev-CADE-2003,UHustadtBKonev-CollegiumLogicum-2004,trppptool}.
The production rules of \trp are shown in Tab.~\ref{tab:productionrules}. The first column assigns a name to a production rule. The second and fourth columns list the premises. The sixth column gives the conclusion. Columns 3, 5, and 7 are described below. Columns 8--12 become relevant only in later sections.
\begin{table}[t]
\centering
{\tiny
\begin{tabular}{||@{\hspace{0.2em}}p{0.12\linewidth}@{\hspace{0.2em}}||@{\hspace{0.2em}}p{0.12\linewidth}@{\hspace{0.2em}}|@{\hspace{0.2em}}c@{\hspace{0.2em}}|@{\hspace{0.2em}}p{0.13\linewidth}@{\hspace{0.2em}}|@{\hspace{0.2em}}c@{\hspace{0.2em}}|@{\hspace{0.2em}}p{0.29\linewidth}@{\hspace{0.2em}}|@{\hspace{0.2em}}c@{\hspace{0.2em}}||@{\hspace{0.2em}}c@{\hspace{0.2em}}|@{\hspace{0.2em}}c@{\hspace{0.2em}}|@{\hspace{0.2em}}c@{\hspace{0.2em}}|@{\hspace{0.2em}}c@{\hspace{0.2em}}|@{\hspace{0.2em}}c@{\hspace{0.2em}}||}
\hline
\hline
rule &
premise 1 &
part. &
premise 2 &
part. &
conclusion &
part. &
p.1 -- c&
t.s.~1&
p.2 -- c&
t.s.~2&
vt. c
\\
\hline
\hline
\multicolumn{12}{||c||}{saturation}
\\
\hline
\hline
init-ii &
\ficlause{\fbor{\dap}{\apres}} &
$\mainpartition$ &
\ficlause{\fbor{\fbnot{\apres}}{\dapp}} &
$\mainpartition$ &
\ficlause{\fbor{\dap}{\dapp}} &
$\mainpartition$ &
\yes &
\no &
\yes &
\no &
\yes
\\
\hline
init-in &
\ficlause{\fbor{\dap}{\apres}} &
$\mainpartition$ &
\fgnclause{\fbor{\fbnot{\apres}}{\dapp}} &
$\mainpartition$ &
\ficlause{\fbor{\dap}{\dapp}} &
$\mainpartition$ &
\yes &
\no &
\yes &
\no &
\yes
\\
\hline
step-nn &
\fgnclause{\fbor{\dap}{\apres}} &
$\mainpartition$ &
\fgnclause{\fbor{\fbnot{\apres}}{\dapp}} &
$\mainpartition$ &
\fgnclause{\fbor{\dap}{\dapp}} &
$\mainpartition$ &
\yes &
\no &
\yes &
\no &
\yes
\\
\hline
step-nx &
\fgnclause{\fbor{\dap}{\apres}} &
$\mainpartition$ &
\fgnxclause{\dapp}{\fbor{\fbnot{\apres}}{\dappp}} &
$\mainpartition$ &
\fgnxclause{\dapp}{\fbor{\dap}{\dappp}} &
$\mainpartition$ &
\yes &
\yes &
\yes &
\no &
\yes
\\
\hline
step-xx &
\fgnxclause{\dap}{\fbor{\dapp}{\apres}} &
$\mainorlooppartition$ &
\fgnxclause{\dappp}{\fbor{\fbnot{\apres}}{\dapppp}} &
$\mainorlooppartition$ &
\fgnxclause{\fbor{\dap}{\dappp}}{\fbor{\dapp}{\dapppp}} &
$\mainorlooppartition$ &
\yes &
\no &
\yes &
\no &
\yes
\\
\hline
\hline
\multicolumn{12}{||c||}{augmentation}
\\
\hline
\hline
aug1 &
\multicolumn{3}{@{\hspace{0.2em}}l@{\hspace{0.2em}}|@{\hspace{0.2em}}}{\feclause{\dap}{\apres}} &
$\mainpartition$ &
\fgnclause{\fbor{\dap}{\fbor{\apres}{\fapwaitfor{\apres}}}} &
$\mainpartition$ &
\yes &
\no &
\na &
\na &
\yes
\\
\hline
aug2 &
\multicolumn{3}{@{\hspace{0.2em}}l@{\hspace{0.2em}}|@{\hspace{0.2em}}}{\feclause{\dap}{\apres}} &
$\mainpartition$ &
\fgnxclause{\fbnot{\fapwaitfor{\apres}}}{\fbor{\apres}{\fapwaitfor{\apres}}} &
$\mainpartition$ &
\no &
\no &
\na &
\na &
\yes
\\
\hline
\hline
\multicolumn{12}{||c||}{BFS loop search}
\\
\hline
\hline
BFS-loop-it-init-x &
\multicolumn{3}{@{\hspace{0.2em}}l@{\hspace{0.2em}}|@{\hspace{0.2em}}}{$\clause \definedas \fgnxclause{\dap}{\dapp}$ with $\fcardinality{\dapp} > 0$} &
$\mainpartition$ &
\clause &
$\looppartition$ &
\yes &
\no &
\na &
\na &
\yes
\\
\hline
BFS-loop-it-init-n &
\multicolumn{3}{@{\hspace{0.2em}}l@{\hspace{0.2em}}|@{\hspace{0.2em}}}{\fgnclause{\dap}} &
$\mainpartition$ &
\fgnxclause{\false}{\dap} &
$\looppartition$ &
\yes &
\yes &
\na &
\na &
\yes
\\
\hline
BFS-loop-it-init-c &
\fgnclause{\dap} &
$\looppartitionp$ &
\feclause{\dapp}{\apres} &
$\mainpartition$ &
\fgnxclause{\false}{\fbor{\dap}{\apres}} &
$\looppartition$ &
\no &
\no &
\no &
\no &
\yes
\\
\hline
BFS-loop-it-sub &
\multicolumn{3}{@{\hspace{0.2em}}l@{\hspace{0.2em}}|@{\hspace{0.2em}}}{\clause \definedas \fgnclause{\dap} \mbox{ with } \fbimplies{\clause}{\fgnclause{\dapp}}} &
$\looppartition$ &
\fgnxclause{\false}{\fbor{\dapp}{\apres}} generated by BFS-loop-it-init-c &
$\looppartition$ &
\yes &
\yes &
\na &
\na &
\no 
\\
\hline
BFS-loop-conclusion1 &
\fgnclause{\dap} &
$\looppartition$ &
\feclause{\dapp}{\apres} &
$\mainpartition$ &
\fgnclause{\fbor{\dap}{\fbor{\dapp}{\apres}}} &
$\mainpartition$ &
\yes &
\no &
\yes &
\no &
\yes
\\
\hline
BFS-loop-conclusion2 &
\fgnclause{\dap} &
$\looppartition$ &
\feclause{\dapp}{\apres} &
$\mainpartition$ &
\fgnxclause{\fbnot{\fapwaitfor{\apres}}}{\fbor{\dap}{\apres}} &
$\mainpartition$ &
\yes &
\yes &
\no &
\no &
\yes
\\
\hline
\hline
\end{tabular}
}
\caption{\label{tab:productionrules}Production rules in \trp. Let $\dap \definedas \bigvee_{\pos = 1 \ldots \maxind} \ap_\pos$, $\dapp \definedas \bigvee_{\pos = 1 \ldots \maxindp} \app_\pos$, $\dappp \definedas \bigvee_{\pos = 1 \ldots \maxindpp} \appp_\pos$, $\dapppp \definedas  \bigvee_{\pos = 1 \ldots \maxindppp} \apppp_\pos$.}
\end{table}

{
\def\checkterm{\lIf{$\emptyclause \in \mainpartition$}{\Return{\unsat}}}
\begin{algorithm}[t]
\SetKw{Or}{or}
\scriptsize
\KwIn{A set of SNF clauses $\setclauses$. \hspace{1em}{\bf Output:} $\Unsat$ if $\setclauses$ is unsatisfiable; $\sat$ otherwise.}
\BlankLine
\algassign{\mainpartition}{\setclauses}; \checkterm\;\nllabel{line:alg:ltlsattrp:init1}
saturate(\mainpartition); \checkterm\;\nllabel{line:alg:ltlsattrp:sat1}
augment(\mainpartition)\;\nllabel{line:alg:ltlsattrp:aug}
saturate(\mainpartition); \checkterm\;\nllabel{line:alg:ltlsattrp:sat2}
\algassign{\mainpartitionp}{\emptyset}\;
\While{$\mainpartitionp \ne \mainpartition$}{
  \algassign{\mainpartitionp}{\mainpartition}\;
  \For{$\clause \in \setclauses \;.\; \clause \mbox{ is an eventuality clause}$}{
    \algassign{\setclausesp}{\{\emptyclause\}}\;\nllabel{line:alg:ltlsattrp:loopsearchinit}
    \Repeat{found \Or $\setclausesp = \emptyset$}{
      initialize-BFS-loop-search-iteration(\mainpartition, \clause, \setclausesp, \looppartition)\;\nllabel{line:alg:ltlsattrp:loopitinit}
      saturate-step-xx(\looppartition)\;\nllabel{line:alg:ltlsattrp:loopitsat}
      \algassign{\setclausesp}{\{\clausep \in \looppartition \mid \clausep \mbox{ has empty $\fnextname$ part}\}}\;\nllabel{line:alg:ltlsattrp:loopitchecksub1}
      \algassign{\setclausespp}{\{\fgnclause{\dapp} \mid \fgnxclause{\false}{\fbor{\dapp}{\apres}} \in \looppartition \mbox{ generated by {\tiny{\setlength{\fboxsep}{1pt}\fbox{BFS-loop-it-init-c}}}}\}}\;\nllabel{line:alg:ltlsattrp:loopitchecksub2}
      \algassign{found}{subsumes(\setclausesp, \setclausespp)}\;\nllabel{line:alg:ltlsattrp:loopitchecksub3}\nllabel{line:alg:ltlsattrp:loopitend}
    }\nllabel{line:alg:ltlsattrp:loopsearchend1}
    \If{found}{
      derive-BFS-loop-search-conclusions(\clause, \setclausesp, \mainpartition)\;\nllabel{line:alg:ltlsattrp:loopsearchconclusions}
      saturate(\mainpartition); \checkterm\;\nllabel{line:alg:ltlsattrp:sat3}
    }\nllabel{line:alg:ltlsattrp:loopsearchend2}
  }
}
\Return{\sat}\;\nllabel{line:alg:ltlsattrp:returnsat}
\caption{\scriptsize\label{alg:ltlsattrp}LTL satisfiability checking via \tr in \trp.}
\end{algorithm}
}

Algorithm \ref{alg:ltlsattrp} provides a high level view of \tr in \trp \cite{UHustadtBKonev-CollegiumLogicum-2004}. The algorithm takes a set of starting clauses $\setclauses$ in SNF as input. It returns $\unsat$ if $\setclauses$ is found to be unsatisfiable (by deriving $\emptyclause$) and $\sat$ otherwise.
Resolution between two initial or two global clauses or between an initial and a global clause is performed by a simple extension of propositional resolution\ifqaplsubmission\else\xspace(e.g., \cite{JRobinson-JACM-1965,JFrancoJMartin-HandbookOfSatisfiability-2009,LBachmairHGanzinger-HandbookOfAutomatedReasoning-2001})\xspace\fi. The corresponding production rules are listed under \emph{saturation} in Tab.~\ref{tab:productionrules}. Given a set of SNF clauses $\setclauses$ we say that one \emph{saturates} $\setclauses$ if one applies these production rules to clauses in $\setclauses$ until no new clauses are generated.
Resolution between a set of initial and global clauses and an eventuality clause with eventuality literal $\apres$ requires finding a set of global clauses that allows to infer conditions under which $\fnext{\fglobally{\fbnot{\apres}}}$ holds. Such a set of clauses is called a \emph{loop} in $\fbnot{\apres}$.
Loop search involves all production rules in Tab.~\ref{tab:productionrules} except \refinitii, \refinitin, \refstepnn, and \refstepnx.

In line \ref{line:alg:ltlsattrp:init1} Alg.~\ref{alg:ltlsattrp} initializes $\mainpartition$ with the set of starting clauses and terminates iff one of these is the empty clause. Then, in line \ref{line:alg:ltlsattrp:sat1}, it saturates $\mainpartition$ (terminating iff the empty clause is generated). In line \ref{line:alg:ltlsattrp:aug} it \emph{augments} $\mainpartition$ by applying production rule \refaugone to each eventuality clause in $\mainpartition$ and \refaugtwo once per eventuality literal in $\mainpartition$, where $\fapwaitfor{\apres}$ is a fresh proposition. This is followed by another round of saturation in line \ref{line:alg:ltlsattrp:sat2}.
From now on Alg.~\ref{alg:ltlsattrp} alternates between searching for a loop for some eventuality clause $\clause$ (lines \ref{line:alg:ltlsattrp:loopsearchinit}--\ref{line:alg:ltlsattrp:loopsearchconclusions}) and saturating $\mainpartition$ if loop search has generated new clauses (line \ref{line:alg:ltlsattrp:sat3}).
It terminates if either the empty clause was derived (line \ref{line:alg:ltlsattrp:sat3}) or if no new clauses were generated (line \ref{line:alg:ltlsattrp:returnsat}).

Loop search for some eventuality clause $\clause$ may take several \emph{iterations} (lines \ref{line:alg:ltlsattrp:loopitinit}--\ref{line:alg:ltlsattrp:loopitend}). Each loop search iteration uses saturation restricted to \refstepxx as a subroutine (line \ref{line:alg:ltlsattrp:loopitsat}). Therefore, each loop search iteration has its own set of clauses $\looppartition$ in which it works. We call $\mainpartition$ and $\looppartition$ \emph{partitions}. Columns 3, 5, and 7 in Tab.~\ref{tab:productionrules} indicate whether a premise (resp.~conclusion) of a production rule is taken from (resp.~put into) the main partition ($\mainpartition$), the loop partition of the current loop search iteration ($\looppartition$), the loop partition of the previous loop search iteration ($\looppartitionp$), or either of $\mainpartition$ or $\looppartition$ as long as premises and conclusion are in the same partition ($\mainorlooppartition$). In line \ref{line:alg:ltlsattrp:loopitinit} partition $\looppartition$ of a loop search iteration is initialized by applying production rule \refloopitinitx once for each global clause with non-empty $\fnextname$ part in $\mainpartition$, rule \refloopitinitn once for each global clause with empty $\fnextname$ part in $\mainpartition$, and rule \refloopitinitc once for each global clause with empty $\fnextname$ part in the partition of the previous loop search iteration $\looppartitionp$. Notice that by construction at this point $\looppartition$ contains only global clauses with non-empty $\fnextname$ part. Then $\looppartition$ is saturated using only rule \refstepxx (line \ref{line:alg:ltlsattrp:loopitsat}). A loop has been found iff each global clause with empty $\fnextname$ part that was derived in the previous loop search iteration is subsumed by at least one global clause with empty $\fnextname$ part that was derived in the current loop search iteration (lines \ref{line:alg:ltlsattrp:loopitchecksub1}--\ref{line:alg:ltlsattrp:loopitchecksub3}). Subsumption between a pair of clauses corresponds to an instance of production rule \refloopitsub; note, though, that this rule does not produce a new clause but records a relation between two clauses to be used later for extraction of a \uc. Loop search for $\clause$ terminates, if either a loop has been found or no clauses with empty $\fnextname$ part were derived (line \ref{line:alg:ltlsattrp:loopsearchend1}). If a loop has been found, rules \refloopconclusionone and \refloopconclusiontwo are applied once to each global clause with empty $\fnextname$ part that was derived in the current loop search iteration (line \ref{line:alg:ltlsattrp:loopsearchconclusions}) to obtain the loop search conclusions for the main partition.

\section{\UC Extraction via \TR}
\label{sec:coreextraction}

In this section we restate the main definitions from \cite{VSchuppan-NFM-2013-full-arXiv} that show how to construct \ucs via \tr.
\ifqaplsubmission
\else
In Sec.~\ref{sec:coreextractioni} we extend this construction to include information on when parts of a \uc are relevant for unsatisfiability.
\fi
Then we present an example for extraction of a \uc in SNF, which is extended with sets of time points in Sec.~\ref{sec:coreextractioni}.

\ifqaplsubmission
\else
\paragraph{Extraction of \UCs in SNF}
\fi

Given an unsatisfiable set of SNF clauses $\setclauses$ we would first like to obtain a subset of $\setclauses$, $\setclausesuc$, that is by itself unsatisfiable from an execution of Alg.~\ref{alg:ltlsattrp}.
The general idea of the construction is unsurprising in that during the execution of Alg.~\ref{alg:ltlsattrp} a resolution graph is built that records which clauses were used to generate other clauses (Def.~\ref{def:resolutiongraph}).
Then the resolution graph is traversed backwards from the empty clause to find the subset of $\setclauses$ that was actually used to prove unsatisfiability (Def.~\ref{def:coreextraction}).
The main concern of Def.~\ref{def:resolutiongraph}, \ref{def:coreextraction}, and their proof of correctness in Thm.~\ref{thm:coreisunsat} (see \cite{VSchuppan-NFM-2013-full-arXiv}) is therefore that certain parts of the \tr proof do not need to be taken into account when determining $\setclausesuc$.

\mydefinition{Resolution Graph}\label{def:resolutiongraph}\cite{VSchuppan-NFM-2013-full-arXiv}
A \emph{resolution graph} $\graph$ is a directed graph consisting of
\begin{inparaenum}
\item a set of vertices $\setvertices$,
\item a set of directed edges $\setedges \subseteq \setvertices \times \setvertices$,
\item a labeling of vertices with SNF clauses $\vertexlabelingname : \setvertices \rightarrow \allclauses$, and
\item a partitioning $\partitioningv$ of the set of vertices $\setvertices$ into one main partition $\mainpartitionv$ and one partition $\looppartitionv_\pos$ for each BFS loop search iteration: $\partitioningv : \setvertices = \fdisjunion{\mainpartitionv}{\fdisjunion{\looppartitionv_0}{\fdisjunion{\ldots}{\looppartitionv_\maxind}}}$.\footnote{$\fdisjunionname$ denotes disjoint union of sets.}
\end{inparaenum}
Let $\setclauses$ be a set of SNF clauses. During an execution of Alg.~\ref{alg:ltlsattrp} with input $\setclauses$ a resolution graph $\graph$ is constructed as follows.

In line \ref{line:alg:ltlsattrp:init1} $\graph$ is initialized:
\begin{inparaenum}
\item $\setvertices$ contains one vertex $\vertex$ per clause $\clause$ in $\setclauses$: $\setvertices = \{\vertex_\clause \mid \clause \in \setclauses\}$,
\item $\setedges$ is empty: $\setedges = \emptyset$,
\item each vertex is labeled with the corresponding clause: $\vertexlabelingname : \setvertices \rightarrow \setclauses, \fvertexlabeling{\vertex_\clause} = \clause$, and
\item the partitioning $\partitioningv$ contains only the main partition $\mainpartitionv$, which contains all vertices: $\partitioningv : \mainpartitionv = \setvertices$.
\end{inparaenum}

Whenever a new BFS loop search iteration is entered (line \ref{line:alg:ltlsattrp:loopitinit}), a new partition $\looppartitionv_\pos$ is created and added to $\partitioningv$.
For each application of a production rule from Tab.~\ref{tab:productionrules} that either generates a new clause in partition $\mainpartition$ or $\looppartition$ or is the first application of rule \refloopitsub to clause $\clausepp$ in $\setclausespp$ in line \ref{line:alg:ltlsattrp:loopitchecksub3}:
\begin{inparaenum}
\item if column 12 (\emph{vt.~c}) of Tab.~\ref{tab:productionrules} contains \yes, then a new vertex $\vertex$ is created for the conclusion $\clause$ (which is a new clause), labeled with $\clause$, and put into partition $\mainpartitionv$ or $\looppartitionv_\pos$;
\item if column 8 (\emph{p.1 -- c}) (resp.~column 10 (\emph{p.2 -- c})) contains \yes, then an edge is created from the vertex labeled with premise 1 (resp.~premise 2) in partition $\mainpartitionv$ or $\looppartitionv_\pos$ to the vertex labeled with the conclusion in partition $\mainpartitionv$ or $\looppartitionv_\pos$.
\end{inparaenum}
\end{definition}

\mydefinition{\UC in SNF}\label{def:coreextraction}\cite{VSchuppan-NFM-2013-full-arXiv}
Let $\setclauses$ be a set of SNF clauses to which Alg.~\ref{alg:ltlsattrp} has been applied and shown unsatisfiability, let $\graph$ be the resolution graph, and let $\vertex_\emptyclause$ be the (unique) vertex in the main partition $\mainpartitionv$ of the resolution graph $\graph$ labeled with the empty clause $\emptyclause$.
Let $\graphp$ be the smallest subgraph of $\graph$ that contains $\vertex_\emptyclause$ and all vertices in $\graph$ (and the corresponding edges) that are backward reachable from $\vertex_\emptyclause$.
The \emph{\uc of $\setclauses$ in SNF}, $\setclausesuc$, is the subset of $\setclauses$ such that there exists a vertex $\vertex$ in the subgraph $\graphp$, labeled with $\clause \in \setclauses$, and contained in the main partition $\mainpartitionv$ of $\graph$: $\setclausesuc = \{\clause \in \setclauses \mid \exists \vertex \in \setvertices_\graphp \;.\; \fband{\fvertexlabeling{\vertex} = \clause}{\vertex \in \mainpartitionv}\}$.
\end{definition}

\mytheorem%
{\thmcoreisunsat}%
{thm:coreisunsat}%
{Unsatisfiability of \UC in SNF}%
{\cite{VSchuppan-NFM-2013-full-arXiv} Let $\setclauses$ be a set of SNF clauses to which Alg.~\ref{alg:ltlsattrp} has been applied and shown unsatisfiability, and let $\setclausesuc$ be the \uc of $\setclauses$ in SNF. Then $\setclausesuc$ is unsatisfiable.}
\thmcoreisunsat{true}

\ifqaplsubmission
\else
\paragraph{From LTL to SNF and Back}
\fi

We now lift the extraction of \ucs from SNF to LTL by restating the translation from LTL to SNF and the mapping from a \uc in SNF back to LTL from \cite{VSchuppan-NFM-2013-full-arXiv}.

\mydefinition{Translation from LTL to SNF}\label{def:ltltosnf}\cite{VSchuppan-NFM-2013-full-arXiv}
Let $\inp$ be an LTL formula over atomic propositions $\allaps$, and let $\alldCNFvars = \{\dCNFvar, \dCNFvarp, \ldots\}$ be a set of fresh atomic propositions not in $\allaps$. Assign each occurrence of a subformula $\prt$ in $\inp$ a Boolean value or a proposition according to col.~2 of Tab.~\ref{tab:ltltosnf}, which is used to reference $\prt$ in the SNF clauses for its superformula. Moreover, assign each occurrence of $\prt$ a set of SNF clauses according to col.~3 or 4 of Tab.~\ref{tab:ltltosnf}. Let $\fdCNFaux{\inp}$ be the set of all SNF clauses obtained from $\inp$ that way. Then the \emph{SNF of $\inp$} is defined as $\fdCNF{\inp} \definedas \fband{\fdCNFvar{\inp}}{\bigwedge_{\dCNFconj \in \fdCNFaux{\inp}} \dCNFconj}$.

\begin{table}[t]
\centering
{\tiny
\begin{tabular}{|l|l|l|l|}
\hline
\hline
Subf. & Prop. & SNF Clauses (positive polarity occurrences) & SNF Clauses (negative polarity occurrences) \\
\hline
\hline
$\true$/$\false$/$\ap$ & $\true$/$\false$/$\ap$ & --- & --- \\
\hline
$\fbnot{\prt}$ & $\fdCNFvar{\fbnot{\prt}}$ & $\fgnclause{\fbimplies{\fdCNFvar{\fbnot{\prt}}}{(\fbnot{\fdCNFvarm{\prt}})}}$ & $\fgnclause{\fbimplies{(\fbnot{\fdCNFvar{\fbnot{\prt}}})}{\fdCNFvarm{\prt}}}$ \\
\hline
$\fband{\prt}{\prtp}$ & $\fdCNFvar{\fband{\prt}{\prtp}}$ & $\fgnclause{\fbimplies{\fdCNFvar{\fband{\prt}{\prtp}}}{\fdCNFvarm{\prt}}}$, $\fgnclause{\fbimplies{\fdCNFvar{\fband{\prt}{\prtp}}}{\fdCNFvarm{\prtp}}}$ & $\fgnclause{\fbimplies{(\fbnot{\fdCNFvar{\fband{\prt}{\prtp}}})}{(\fbor{(\fbnot{\fdCNFvarm{\prt}})}{(\fbnot{\fdCNFvarm{\prtp}})})}}$ \\
\hline
$\fbor{\prt}{\prtp}$ & $\fdCNFvar{\fbor{\prt}{\prtp}}$ & $\fgnclause{\fbimplies{\fdCNFvar{\fbor{\prt}{\prtp}}}{(\fbor{\fdCNFvarm{\prt}}{\fdCNFvarm{\prtp}})}}$ & $\fgnclause{\fbimplies{(\fbnot{\fdCNFvar{\fbor{\prt}{\prtp}}})}{(\fbnot{\fdCNFvarm{\prt}})}}$, $\fgnclause{\fbimplies{(\fbnot{\fdCNFvar{\fbor{\prt}{\prtp}}})}{(\fbnot{\fdCNFvarm{\prtp}})}}$ \\
\hline
$\fnext{\prt}$ & $\fdCNFvar{\fnext{\prt}}$ & $\fgnclause{\fbimplies{\fdCNFvar{\fnext{\prt}}}{(\fnext{\fdCNFvarm{\prt}})}}$ & $\fgnclause{\fbimplies{(\fbnot{\fdCNFvar{\fnext{\prt}}})}{(\fnext{\fbnot{\fdCNFvarm{\prt}}})}}$ \\
\hline
$\fglobally{\prt}$ & $\fdCNFvar{\fglobally{\prt}}$ & $\fgnclause{\fbimplies{\fdCNFvar{\fglobally{\prt}}}{(\fnext{\fdCNFvar{\fglobally{\prt}}})}}$, $\fgnclause{\fbimplies{\fdCNFvar{\fglobally{\prt}}}{\fdCNFvarm{\prt}}}$ & $\fgnclause{\fbimplies{(\fbnot{\fdCNFvar{\fglobally{\prt}}})}{(\ffinally{\fbnot{\fdCNFvarm{\prt}}})}}$ \\
\hline
$\ffinally{\prt}$ & $\fdCNFvar{\ffinally{\prt}}$ & $\fgnclause{\fbimplies{\fdCNFvar{\ffinally{\prt}}}{(\ffinally{\fdCNFvarm{\prt}})}}$ & $\fgnclause{\fbimplies{(\fbnot{\fdCNFvar{\ffinally{\prt}}})}{(\fnext{\fbnot{\fdCNFvar{\ffinally{\prt}}}})}}$, $\fgnclause{\fbimplies{(\fbnot{\fdCNFvar{\ffinally{\prt}}})}{(\fbnot{\fdCNFvarm{\prt}})}}$ \\
\hline
$\funtil{\prt}{\prtp}$ & $\fdCNFvar{\funtil{\prt}{\prtp}}$ & $\fgnclause{\fbimplies{\fdCNFvar{\funtil{\prt}{\prtp}}}{(\fbor{\fdCNFvarm{\prtp}}{\fdCNFvarm{\prt}})}}$, $\fgnclause{\fbimplies{\fdCNFvar{\funtil{\prt}{\prtp}}}{(\fbor{\fdCNFvarm{\prtp}}{(\fnext{\fdCNFvar{\funtil{\prt}{\prtp}}})})}}$, & $\fgnclause{\fbimplies{(\fbnot{\fdCNFvar{\funtil{\prt}{\prtp}}})}{(\fbnot{\fdCNFvarm{\prtp}})}}$, $\fgnclause{\fbimplies{(\fbnot{\fdCNFvar{\funtil{\prt}{\prtp}}})}{(\fbor{(\fbnot{\fdCNFvarm{\prt}})}{(\fnext{\fbnot{\fdCNFvar{\funtil{\prt}{\prtp}}}})})}}$ \\
& & $\fgnclause{\fbimplies{\fdCNFvar{\funtil{\prt}{\prtp}}}{(\ffinally{\fdCNFvarm{\prtp}})}}$ & \\
\hline
$\freleases{\prt}{\prtp}$ & $\fdCNFvar{\freleases{\prt}{\prtp}}$ & $\fgnclause{\fbimplies{\fdCNFvar{\freleases{\prt}{\prtp}}}{\fdCNFvarm{\prtp}}}$, $\fgnclause{\fbimplies{\fdCNFvar{\freleases{\prt}{\prtp}}}{(\fbor{\fdCNFvarm{\prt}}{(\fnext{\fdCNFvar{\freleases{\prt}{\prtp}}})})}}$ & $\fgnclause{\fbimplies{(\fbnot{\fdCNFvar{\freleases{\prt}{\prtp}}})}{(\fbor{(\fbnot{\fdCNFvarm{\prtp}})}{(\fbnot{\fdCNFvarm{\prt}})})}}$, $\fgnclause{\fbimplies{(\fbnot{\fdCNFvar{\freleases{\prt}{\prtp}}})}{(\fbor{(\fbnot{\fdCNFvarm{\prtp}})}{(\fnext{\fbnot{\fdCNFvar{\freleases{\prt}{\prtp}}}})})}}$, \\
& & & $\fgnclause{\fbimplies{(\fbnot{\fdCNFvar{\freleases{\prt}{\prtp}}})}{(\ffinally{\fbnot{\fdCNFvarm{\prtp}}})}}$ \\
\hline
\hline
\end{tabular}
}
\caption{\label{tab:ltltosnf}Translation from LTL to SNF.}
\end{table}
\end{definition}

\mydefinition{Mapping a \UC in SNF to a \UC in LTL}\label{def:ltlcoreextraction}\cite{VSchuppan-NFM-2013-full-arXiv}
Let $\inp$ be an unsatisfiable LTL formula, let $\fdCNF{\inp}$ be its SNF, and let $\setclausesuc$ be the \uc of $\fdCNF{\inp}$ in SNF. Then the \emph{\uc of $\inp$ in LTL}, $\inpuc$, is obtained as follows. For each positive (resp.~negative) polarity occurrence of a proper subformula $\prt$ of $\inp$ with proposition $\fdCNFvar{\prt}$ according to Tab.~\ref{tab:ltltosnf}, replace $\prt$ in $\inp$ with $\true$ (resp.~$\false$) iff $\setclausesuc$ contains no clause with an occurrence of proposition $\fdCNFvar{\prt}$ that is marked {\color{blue}\setlength\fboxsep{1pt}\fbox{blue boxed}} in Tab.~\ref{tab:ltltosnf}. (We are sloppy in that we ``replace'' subformulas of replaced subformulas, while in effect they simply vanish.)
\end{definition}

\mytheorem%
{\thmltlcoreisunsat}%
{thm:ltlcoreisunsat}%
{Unsatisfiability of \UC in LTL}%
{\cite{VSchuppan-NFM-2013-full-arXiv} Let $\inp$ be an unsatisfiable LTL formula, and let $\inpuc$ be the \uc of $\inp$ in LTL. Then $\inpuc$ is unsatisfiable.}
\thmltlcoreisunsat{true}

\ifqaplsubmission
\else
\paragraph{Example}
\fi

In Fig.~\ref{fig:exresgraphucslsex} we show an example of an execution of the \tr algorithm with the corresponding resolution graph and \uc extraction in SNF.
The set of SNF clauses $\setclauses$ to be solved contains $a$, $\fglobally{(\fbor{(\fbnot{a})}{\fnext{b}})}$, $\fglobally{(\fbor{(\fbnot{b})}{\fnext{a}})}$, $\fglobally{(\fbor{(\fbnot{a})}{\fbnot{c}})}$, $\fglobally{(\fbor{(\fbnot{c})}{\fnext{\fbnot{a}}})}$, and $\fglobally{(\ffinally{c})}$.
The first three clauses $a$, $\fglobally{(\fbor{(\fbnot{a})}{\fnext{b}})}$, and $\fglobally{(\fbor{(\fbnot{b})}{\fnext{a}})}$ force $a$ to be $\true$ at even time points.
This is contradicted by the last three clauses $\fglobally{(\fbor{(\fbnot{a})}{\fbnot{c}})}$, $\fglobally{(\fbor{(\fbnot{c})}{\fnext{\fbnot{a}}})}$, and $\fglobally{(\ffinally{c})}$: they require that $a$ eventually becomes $\false$ for two consecutive time points.
Clearly, $\setclauses$ is unsatisfiable.
This example is based on the same idea as \eqref{ex:everysecond} in Sec.~\ref{sec:ltli}.
However, the SNF obtained by our translation from LTL to SNF for \eqref{ex:everysecond} is larger than $\setclauses$, with the corresponding figure harder to fit on one page.

In Fig.~\ref{fig:exresgraphucslsex} the \tr algorithm proceeds from bottom to top.
Clauses are connected with edges according to cols.~8 and 10 of Tab.~\ref{tab:productionrules} and labeled with the corresponding production rules, where ``BFS-loop'' is abbreviated to ``loop'', ``init'' to ``i'', and ``conclusion'' to ``conc''.
In the first row from the bottom (in the light red shaded rectangle) are the starting clauses from $\setclauses$.
In the top right corner is the empty clause $\emptyclause$ signaling unsatisfiability of $\setclauses$.
Row 2 contains the clauses resulting from the first round of saturation (line \ref{line:alg:ltlsattrp:sat1} in Alg.~\ref{alg:ltlsattrp}) and from augmentation (line \ref{line:alg:ltlsattrp:aug}).\footnote{While it may seem that some clauses are not considered for loop initialization or saturation, this is due to either subsumption of one clause by another (e.g., $\fglobally{(\fbor{(\fbnot{\fapwaitfor{c}})}{\fnext{(\fbor{c}{\fapwaitfor{c}})}})}$ by $\fglobally{(\fbor{c}{\fapwaitfor{c}})}$) or the fact that \trp uses \emph{ordered} resolution (e.g., $a$ with $\fglobally{(\fbor{(\fbnot{a})}{\fbnot{c}})}$; \ifqaplsubmission\cite{UHustadtBKonev-CADE-2003}\else\cite{UHustadtBKonev-CADE-2003,LBachmairHGanzinger-HandbookOfAutomatedReasoning-2001}\fi). Both are issues of completeness of \tr and, therefore, not discussed in this paper.}
The second round of saturation (line \ref{line:alg:ltlsattrp:sat2}) produces no new clauses.
The dark green shaded rectangle is the partition for the first iteration of a loop search for a loop in $\fbnot{c}$.
Row 3 contains the result of loop search initialization (line \ref{line:alg:ltlsattrp:loopitinit}) and row 4 the clauses obtained by restricted saturation (line \ref{line:alg:ltlsattrp:loopitsat}).
As none of the clauses in row 4 subsumes $\emptyclause$, this iteration terminates without having found a loop.
The second loop search iteration is in the light green shaded rectangle.
Again row 5 contains the result of loop search initialization and row 6 the clauses obtained by restricted saturation.
This time the subsumption test is successful (lines \ref{line:alg:ltlsattrp:loopitchecksub1}--\ref{line:alg:ltlsattrp:loopitend}), and row 7 shows the loop search conclusions (line \ref{line:alg:ltlsattrp:loopsearchconclusions}).
The last row finally contains the derivation of $\emptyclause$ by saturation (line \ref{line:alg:ltlsattrp:sat3}).

The clauses that are backward reachable from $\emptyclause$ are shown in blue with blue, thick, dashed boxes.
The corresponding edges are thick, blue or red, and dashed or dotted.
The resulting \uc comprises all clauses in $\setclauses$ (note that this example shows the mechanism rather than the benefits of extracting \ucs).

The distinction between blue, dashed and red, dotted edges as well as the sets of time points shown in black boxes are needed when sets of time points are added in Sec.~\ref{sec:coreextractioni}.
Please ignore those for now.

{
\newcommand{\refshortinferencerule}[1]{{\tiny{#1}}\xspace}
\newcommand{\refshortinitii}{\refshortinferencerule{init-ii}}
\newcommand{\refshortinitin}{\refshortinferencerule{init-in}}
\newcommand{\refshortstepnn}{\refshortinferencerule{step-nn}}
\newcommand{\refshortstepnx}{\refshortinferencerule{step-nx}}
\newcommand{\refshortstepxx}{\refshortinferencerule{step-xx}}
\newcommand{\refshortloopitinitx}{\refshortinferencerule{loop-it-i-x}}
\newcommand{\refshortloopitinitn}{\refshortinferencerule{loop-it-i-n}}
\newcommand{\refshortloopitinitc}{\refshortinferencerule{loop-it-i-c}}
\newcommand{\refshortloopitsub}{\refshortinferencerule{loop-it-sub}}
\newcommand{\refshortloopconclusionone}{\refshortinferencerule{loop-conc1}}
\newcommand{\refshortloopconclusiontwo}{\refshortinferencerule{loop-conc2}}
\newcommand{\refshortaugone}{\refshortinferencerule{aug1}}
\newcommand{\refshortaugtwo}{\refshortinferencerule{aug2}}

\begin{center}
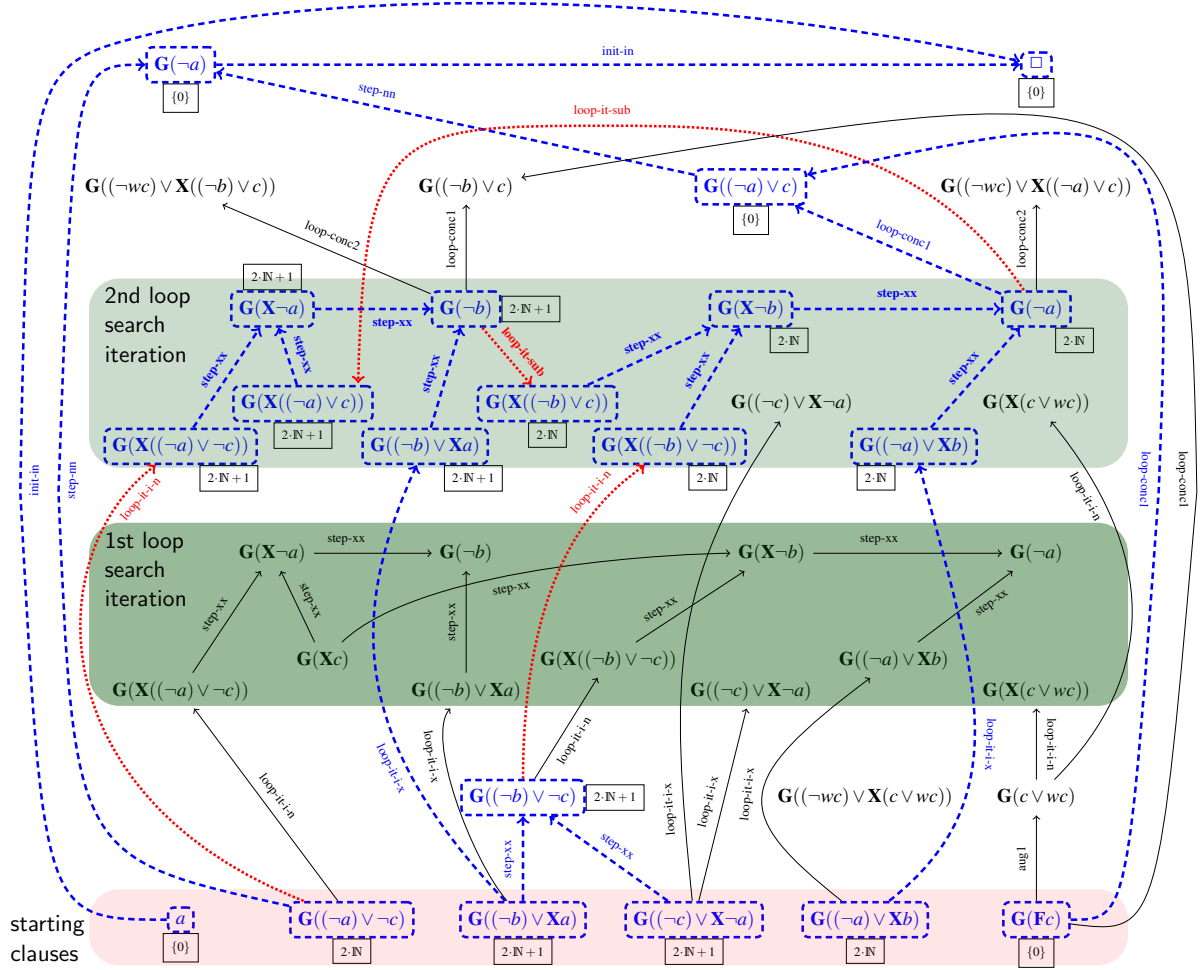
\begin{figure}
\begin{center}
\hspace*{-0em}
\resizebox{\linewidth}{!}{
\begin{tikzpicture}[auto,
    =<triangle 45,
    every edge/.style={draw,sloped,above},
    corev/.style={draw=blue,rectangle,rounded corners=1mm,very thick,densely dashed,text=blue},
    coreea0/.style={blue,very thick,densely dashed},
    coreea1/.style={red,very thick,densely dotted},
    coreel0/.style={draw=none,very thick,text=blue},
    coreel1/.style={draw=none,very thick,text=red},
    bfel/.style={font=\bf},
    sls/.style={draw=,rectangle,font=\tiny},
    scale=1.0]
  \tikzstyle{every node}=[font=\footnotesize]
  \clip (-2.8,-0.8) rectangle (16.7,15.1);
  \node (n_1_0) at (0,0) [corev,label={[sls]below:$\zeroset$}] {$a$};
  \node (n_5_0) at (2.8,0) [corev,label={[sls]below:$\evenset$}] {$\fglobally{(\fbor{(\fbnot{a})}{\fbnot{c}})}$};
  \node (n_3_0) at (5.6,0) [corev,label={[sls]below:$\oddset$}] {$\fglobally{(\fbor{(\fbnot{b})}{\fnext{a}})}$};
  \node (n_6_0) at (8.4,0) [corev,label={[sls]below:$\oddset$}] {$\fglobally{(\fbor{(\fbnot{c})}{\fnext{\fbnot{a}}})}$};
  \node (n_2_0) at (11.2,0) [corev,label={[sls]below:$\evenset$}] {$\fglobally{(\fbor{(\fbnot{a})}{\fnext{b}})}$};
  \node (n_4_0) at (14,0) [corev,label={[sls]below:$\zeroset$}] {$\fglobally{(\ffinally{c})}$};

  \node (n_7_0) at (5.6,2) [corev,label={[sls]right:$\oddset$}] {$\fglobally{(\fbor{(\fbnot{b})}{\fbnot{c}})}$};
  \node (n_8_0) at (11.2,2) {$\fglobally{(\fbor{(\fbnot{\fapwaitfor{c}})}{\fnext{(\fbor{c}{\fapwaitfor{c}})}})}$};
  \node (n_9_0) at (14,2) {$\fglobally{(\fbor{c}{\fapwaitfor{c}})}$};

  \node (n_10_1) at (0,3.75) {$\fglobally{(\fnext{(\fbor{(\fbnot{a})}{\fbnot{c}})})}$};
  \node (n_14_1) at (2.33,4.25)  {$\fglobally{(\fnext{c})}$};
  \node (n_3_1) at (4.67,3.75) {$\fglobally{(\fbor{(\fbnot{b})}{\fnext{a}})}$};
  \node (n_11_1) at (7,4.25) {$\fglobally{(\fnext{(\fbor{(\fbnot{b})}{\fbnot{c}})})}$};
  \node (n_6_1) at (9.33,3.75) {$\fglobally{(\fbor{(\fbnot{c})}{\fnext{\fbnot{a}}})}$};
  \node (n_2_1) at (11.67,4.25) {$\fglobally{(\fbor{(\fbnot{a})}{\fnext{b}})}$};
  \node (n_12_1) at (14,3.75) {$\fglobally{(\fnext{(\fbor{c}{\fapwaitfor{c}})})}$};

  \node (n_15_1) at (1.5,6) {$\fglobally{(\fnext{\fbnot{a}})}$};
  \node (n_17_1) at (4.67,6) {$\fglobally{(\fbnot{b})}$};
  \node (n_16_1) at (9.67,6) {$\fglobally{(\fnext{\fbnot{b}})}$};
  \node (n_18_1) at (14,6) {$\fglobally{(\fbnot{a})}$};

  \node (n_10_2) at (0,7.75) [corev,label={[sls]below right:$\oddset$}] {$\fglobally{(\fnext{(\fbor{(\fbnot{a})}{\fbnot{c}})})}$};
  \node (n_19_2) at (2,8.45) [corev,label={[sls]below:$\oddset$}] {$\fglobally{(\fnext{(\fbor{(\fbnot{a})}{c})})}$};
  \node (n_3_2) at (4,7.75) [corev,label={[sls]below right:$\oddset$}] {$\fglobally{(\fbor{(\fbnot{b})}{\fnext{a}})}$};
  \node (n_20_2) at (6,8.45) [corev,label={[sls]below:$\evenset$}] {$\fglobally{(\fnext{(\fbor{(\fbnot{b})}{c})})}$};
  \node (n_11_2) at (8,7.75) [corev,label={[sls]below right:$\evenset$}] {$\fglobally{(\fnext{(\fbor{(\fbnot{b})}{\fbnot{c}})})}$};
  \node (n_6_2) at (10,8.45) {$\fglobally{(\fbor{(\fbnot{c})}{\fnext{\fbnot{a}}})}$};
  \node (n_2_2) at (12,7.75) [corev,label={[sls]below left:$\evenset$}] {$\fglobally{(\fbor{(\fbnot{a})}{\fnext{b}})}$};
  \node (n_12_2) at (14,8.45) {$\fglobally{(\fnext{(\fbor{c}{\fapwaitfor{c}})})}$};

  \node (n_21_2) at (1.5,10) [corev,label={[sls]above:$\oddset$}] {$\fglobally{(\fnext{\fbnot{a}})}$};
  \node (n_23_2) at (4.67,10) [corev,label={[sls]right:$\oddset$}] {$\fglobally{(\fbnot{b})}$};
  \node (n_24_2) at (9.33,10) [corev,label={[sls]below right:$\evenset$}] {$\fglobally{(\fnext{\fbnot{b}})}$};
  \node (n_25_2) at (14,10) [corev,label={[sls]below right:$\evenset$}] {$\fglobally{(\fbnot{a})}$};

  \node (n_28_0) at (0,12) {$\fglobally{(\fbor{(\fbnot{\fapwaitfor{c}})}{\fnext{(\fbor{(\fbnot{b})}{c})}})}$};
  \node (n_29_0) at (4.67,12) {$\fglobally{(\fbor{(\fbnot{b})}{c})}$};
  \node (n_27_0) at (9.33,12) [corev,label={[sls]below:$\zeroset$}] {$\fglobally{(\fbor{(\fbnot{a})}{c})}$};
  \node (n_26_0) at (14,12) {$\fglobally{(\fbor{(\fbnot{\fapwaitfor{c}})}{\fnext{(\fbor{(\fbnot{a})}{c})}})}$};

  \node (n_30_0) at (0,14) [corev,label={[sls]below:$\zeroset$}] {$\fglobally{(\fbnot{a})}$};
  \node (n_32_0) at (14,14) [corev,label={[sls]below:$\zeroset$}] {$\emptyclause$};

  \fill[fill=red,fill opacity=0.1,rounded corners=5mm] (-1.5,-0.75) rectangle (15.5,0.5);
  \draw (-1.65,-0.3) node [left,text width=1cm,font=\sffamily] {starting \newline clauses};
  \fill[fill=darkgreen,fill opacity=0.4,rounded corners=5mm] (-1.5,3.5) rectangle (15.5,6.5);
  \draw (-1.375,5.75) node [right,text width=2cm,font=\sffamily] {1st loop \newline search \newline iteration};
  \fill[fill=darkgreen,fill opacity=0.2,rounded corners=5mm] (-1.5,7.4) rectangle (15.5,10.5);
  \draw (-1.375,9.75) node [right,text width=2cm,font=\sffamily] {2nd loop \newline search \newline iteration};

  \path[->]
  (n_3_0) edge [coreea0] node [coreel0] {\refshortstepxx} (n_7_0) 
  (n_6_0) edge [coreea0] node [coreel0] {\refshortstepxx} (n_7_0) 
  (n_4_0) edge node {\refshortaugone} (n_9_0) 
  (n_6_0) edge node {\refshortloopitinitx} (n_6_1) 
  (n_5_0) edge node {\refshortloopitinitn} (n_10_1) 
  (n_7_0) edge node [pos=0.45,below] {\refshortloopitinitn} (n_11_1) 
  (n_9_0) edge node [rotate=180] {\refshortloopitinitn} (n_12_1) 
  (n_10_1) edge node {\refshortstepxx} (n_15_1) 
  (n_11_1) edge node [pos=0.3] {\refshortstepxx} (n_16_1) 
  (n_14_1) edge node {\refshortstepxx} (n_15_1) 
  (n_3_1) edge node {\refshortstepxx} (n_17_1) 
  (n_15_1) edge node [pos=0.275] {\refshortstepxx} (n_17_1) 
  (n_2_1) edge node [pos=0.7,below] {\refshortstepxx} (n_18_1) 
  (n_16_1) edge node [pos=0.35] {\refshortstepxx} (n_18_1) 
  (n_10_2) edge [coreea0] node [bfel,coreel0] {\refshortstepxx} (n_21_2) 
  (n_19_2) edge [coreea0] node [bfel,pos=0.45,coreel0] {\refshortstepxx} (n_21_2) 
  (n_3_2) edge [coreea0] node [bfel,coreel0] {\refshortstepxx} (n_23_2) 
  (n_21_2) edge [coreea0] node [bfel,pos=0.67,below,coreel0] {\refshortstepxx} (n_23_2) 
  (n_11_2) edge [coreea0] node [bfel,coreel0] {\refshortstepxx} (n_24_2) 
  (n_20_2) edge [coreea0] node [bfel,coreel0] {\refshortstepxx} (n_24_2) 
  (n_2_2) edge [coreea0] node [bfel,coreel0] {\refshortstepxx} (n_25_2) 
  (n_24_2) edge [coreea0] node [bfel,coreel0] {\refshortstepxx} (n_25_2) 
  (n_23_2) edge [coreea1] node [bfel,pos=0.6,coreel1] {\refshortloopitsub} (n_20_2) 
  (n_25_2) edge node [pos=0.6] {\refshortloopconclusiontwo} (n_26_0) 
  (n_25_2) edge [coreea0] node [coreel0] {\refshortloopconclusionone} (n_27_0) 
  (n_23_2) edge node {\refshortloopconclusiontwo} (n_28_0) 
  (n_23_2) edge node [pos=0.6] {\refshortloopconclusionone} (n_29_0) 
  (n_27_0) edge [coreea0] node [pos=0.67,coreel0] {\refshortstepnn} (n_30_0) 
  (n_30_0) edge [coreea0] node [coreel0] {\refshortinitin} (n_32_0);
  \draw[->] (n_1_0) [coreea0] .. controls (-2.25,-0) .. (-2.625,7) .. controls (-2.5,14) .. node [coreel0,pos=0.01,sloped,below] {\refshortinitin} (-0.5,14.5) .. controls (1.5,15) and (7,15.5) .. (n_32_0);
  \draw[->] (n_5_0) [coreea0] .. controls (-1.75,1) .. (-2,7) .. controls (-1.75,14) .. node [pos=0.01,coreel0,sloped,below] {\refshortstepnn} (n_30_0);
  \draw[->] (n_5_0) [coreea1] .. controls (-2.33,2) and (-2.33,6) .. node [coreel1,pos=0.92,sloped,below] {\refshortloopitinitn} (n_10_2);
  \draw[->] (n_9_0) .. controls (16,4) and (16,6) .. node [pos=0.825,sloped,below] {\refshortloopitinitn} (n_12_2);
  \node (n_19_2_aux) at (2.9,8.65) {}; 
  \draw[->] (n_25_2) [coreea1] .. controls (12,13) and (9,13) .. (7,13) .. controls (3.5,13) and (3,13) .. node [pos=0.01,coreel1,sloped,above] {\refshortloopitsub} (n_19_2_aux);
  \draw[->] (n_4_0) [coreea0] .. controls (15.5,0) .. (16,7) .. controls (16.25,13.5) .. node [coreel0,pos=0.01,sloped,below,rotate=180] {\refshortloopconclusionone} (n_27_0);
  \draw[->] (n_4_0) .. controls (16,-0.25) .. (16.625,7) .. controls (16.5,14) .. node [pos=0.01,sloped,below] {\refshortloopconclusionone} (n_29_0);
  \draw[->] (n_3_0) [coreea0] .. controls (2.75,3) and (2.75,5) .. node [coreel0,pos=0.3,sloped,below] {\refshortloopitinitx} (n_3_2);
  \draw[->] (n_3_0) .. controls (4.75,1) and (4,3) .. node [pos=0.7,sloped,below] {\refshortloopitinitx} (n_3_1);
  \draw[->] (n_14_1) .. controls (4,5.85) and (8.4,6) .. node [pos=0.41,sloped,below] {\refshortstepxx} (n_16_1);
  \draw[->] (n_7_0) [coreea1] .. controls (5.6,6) and (7,7) .. node [coreel1,pos=0.7,sloped,above] {\refshortloopitinitn} (n_11_2);
  \draw[->] (n_6_0) .. controls (8,4) and (8,5.5) .. node [pos=0.15,sloped,above,rotate=180] {\refshortloopitinitx} (n_6_2);
  \draw[->] (n_2_0) [coreea0] .. controls (13.75,2) and (13,4) .. node [coreel0,pos=0.45,sloped,above,rotate=180] {\refshortloopitinitx} (n_2_2);
  \draw[->] (n_2_0) .. controls (9,1.75) and (9,2.25) .. node [sloped,above] {\refshortloopitinitx} (n_2_1);

\end{tikzpicture}
}
\end{center}
\caption{\label{fig:exresgraphucslsex} Example of an execution of the \tr algorithm with corresponding resolution graph and \uc extraction in SNF with sets of time points.}
\end{figure}
\end{center}
}

\section{LTL with Sets of Time Points (\ltli)}
\label{sec:ltli}

In this section we propose a notation that allows to integrate more detailed information from a resolution proof of the unsatisfiability of some LTL formula $\inp$ into the \uc $\inpuc$. The information we are interested in are the time points at which a part of an LTL formula is needed to prove unsatisfiability. Hence, we assign to each subformula a set of time points that indicates at which time points that subformula will be evaluated; at other time points the subformula is considered to be $\true$ or $\false$ depending on polarity. Note that this can be seen as an extension of a notion of \uc in \ifqaplsubmission\cite{VSchuppan-SCP-2012}\else\cite{VSchuppan-SCP-2012,OKupfermanMVardi-STTT-2003}\fi, where subformulas are replaced with $\true$ or $\false$ depending on polarity. We wish to emphasize that it is not our goal to introduce a ``new logic'', but merely to suggest a notation with well defined semantics that allows to smoothly integrate such information.

\mydefinition{\ltli Syntax}
The set of \emph{\ltli formulas} is constructed inductively as follows. The Boolean constants $\false$ (false), $\true$ (true) $\in \allbools$ and any atomic proposition $\ap \in \allaps$ are \ltli formulas. If $\setpos, \setposp \subseteq \allnats$ are sets of time points and if $\prti$, $\prtpi$ are \ltli formulas, so are $\fbnoti{\prti}{\setpos}$ (not), $\fbori{\prti}{\prtpi}{\setpos}{\setposp}$ (or), $\fbandi{\prti}{\prtpi}{\setpos}{\setposp}$ (and), $\fnexti{\prti}{\setpos}$ (next time), $\funtili{\prti}{\prtpi}{\setpos}{\setposp}$ (until), $\freleasesi{\prti}{\prtpi}{\setpos}{\setposp}$ (releases), $\ffinallyi{\prti}{\setpos}$ (finally), and $\fgloballyi{\prti}{\setpos}$ (globally).
$\fbimpliesi{\prti}{\prtpi}{\setpos}{\setposp}$ (implies) abbreviates $\fbori{\fbnoti{\prti}{\setpos}}{\prtpi}{\setpos}{\setposp}$.
\end{definition}

We now recursively define the semantics of an \ltli formula at time points $\pos \in \allnats$ of a word $\iword \in (\mkpowerset{\allaps})^\omega$. Note that the semantics depends on the polarity of the occurrence of a subformula. The intuition for the semantics is that if a time point $\pos$ is not contained in a set $\setpos$, then the corresponding operand at that time point cannot be used to establish unsatisfiability.
\mydefinition{\ltli Semantics}\label{def:ltlisemantics}
\begin{table}[t]
\centering
{\tiny
$
\begin{array}{|@{\hspace{0.5em}}l@{\hspace{1.25em}}c@{\hspace{1.25em}}l@{\hspace{1.25em}}l@{\hspace{0.5em}}|}
\hline
\hline
\mbox{formula} & & \mbox{positive polarity} & \mbox{negative polarity}
\\
\hline
\hline &&&\\[-2ex]
\mkpair{\iword}{\pos} \models \true \mbox{ (resp.~$\false$)} & \proofbiimplies & \true \mbox{ (resp.~$\false$)} & \true \mbox{ (resp.~$\false$)} \\
\hline &&&\\[-2ex]
\mkpair{\iword}{\pos} \models \ap & \proofbiimplies & \ap \in \fiwordgetpos{\iword}{\pos} & \ap \in \fiwordgetpos{\iword}{\pos} \\
\hline &&&\\[-2ex]
\mkpair{\iword}{\pos} \models \fbnoti{\prti}{\setpos} & \proofbiimplies & \fbor{(\pos \not \in \setpos)}{(\mkpair{\iword}{\pos} \not \models \prti)} & \fband{(\pos \in \setpos)}{(\mkpair{\iword}{\pos} \not \models \prti)} \\
\hline &&&\\[-2ex]
\mkpair{\iword}{\pos} \models \fbori{\prti}{\prtpi}{\setpos}{\setposp} & \proofbiimplies & \fbor{(\fbor{(\pos \not \in \setpos)}{(\mkpair{\iword}{\pos} \models \prti)})}{(\fbor{(\pos \not \in \setposp)}{(\mkpair{\iword}{\pos} \models \prtpi)})} & \fbor{(\fband{(\pos \in \setpos)}{(\mkpair{\iword}{\pos} \models \prti)})}{(\fband{(\pos \in \setposp)}{(\mkpair{\iword}{\pos} \models \prtpi)})}\\
\hline &&&\\[-2ex]
\mkpair{\iword}{\pos} \models \fbandi{\prti}{\prtpi}{\setpos}{\setposp} & \proofbiimplies & \fband{(\fbor{(\pos \not \in \setpos)}{(\mkpair{\iword}{\pos} \models \prti)})}{(\fbor{(\pos \not \in \setposp)}{(\mkpair{\iword}{\pos} \models \prtpi)})} & \fband{(\fband{(\pos \in \setpos)}{(\mkpair{\iword}{\pos} \models \prti)})}{(\fband{(\pos \in \setposp)}{(\mkpair{\iword}{\pos} \models \prtpi)})} \\
\hline &&&\\[-2ex]
\mkpair{\iword}{\pos} \models \fnexti{\prti}{\setpos} & \proofbiimplies & \fbor{(\pos + 1 \not \in \setpos)}{(\mkpair{\iword}{\pos + 1} \models \prti)} & \fband{(\pos + 1 \in \setpos)}{(\mkpair{\iword}{\pos + 1} \models \prti)} \\
\hline &&&\\[-2ex]
\mkpair{\iword}{\pos} \models \funtili{\prti}{\prtpi}{\setpos}{\setposp} & \proofbiimplies & \exists \posp \ge \pos \;.\; (\fband{(\fbor{(\posp \not \in \setposp)}{(\mkpair{\iword}{\posp} \models \prtpi)})}{(\forall \pos \le \pospp < \posp \;.\; (\fbor{(\pospp \not \in \setpos)}{(\mkpair{\iword}{\pospp} \models \prti)}))}) & \exists \posp \ge \pos \;.\; (\fband{(\fband{(\posp \in \setposp)}{(\mkpair{\iword}{\posp} \models \prtpi)})}{(\forall \pos \le \pospp < \posp \;.\; (\fband{(\pospp \in \setpos)}{(\mkpair{\iword}{\pospp} \models \prti)}))}) \\
\hline &&&\\[-2ex]
\mkpair{\iword}{\pos} \models \freleasesi{\prti}{\prtpi}{\setpos}{\setposp} & \proofbiimplies & \forall \posp \ge \pos \;.\; (\fbor{(\fbor{(\posp \not \in \setposp)}{(\mkpair{\iword}{\posp} \models \prtpi)})}{(\exists \pos \le \pospp < \posp \;.\; (\fbor{(\pospp \not \in \setpos)}{(\mkpair{\iword}{\pospp} \models \prti)}))}) & \forall \posp \ge \pos \;.\; (\fbor{(\fband{(\posp \in \setposp)}{(\mkpair{\iword}{\posp} \models \prtpi)})}{(\exists \pos \le \pospp < \posp \;.\; (\fband{(\pospp \in \setpos)}{(\mkpair{\iword}{\pospp} \models \prti)}))}) \\
\hline &&&\\[-2ex]
\mkpair{\iword}{\pos} \models \ffinallyi{\prti}{\setpos} & \proofbiimplies & \exists \posp \ge \pos \;.\; (\fbor{(\posp \not \in \setpos)}{(\mkpair{\iword}{\posp} \models \prti)}) & \exists \posp \ge \pos \;.\; (\fband{(\posp \in \setpos)}{(\mkpair{\iword}{\posp} \models \prti)}) \\
\hline &&&\\[-2ex]
\mkpair{\iword}{\pos} \models \fgloballyi{\prti}{\setpos} & \proofbiimplies & \forall \posp \ge \pos \;.\; (\fbor{(\posp \not \in \setpos)}{(\mkpair{\iword}{\posp} \models \prti)}) & \forall \posp \ge \pos \;.\; (\fband{(\posp \in \setpos)}{(\mkpair{\iword}{\posp} \models \prti)}) \\
\hline
\hline
\end{array}
$
}%
\caption{\label{tab:ltlisemantics} Semantics of \ltli. $\iword$ is a word in $(\mkpowerset{\allaps})^\omega$, $\pos$ is a time point in $\allnats$.}
\end{table}
The semantics of \ltli is given in Tab.~\ref{tab:ltlisemantics}. $\iword$ \emph{satisfies} a formula $\inp$ iff the formula holds at the beginning of $\iword$: $\iword \models \inp \proofbiimplies \mkpair{\iword}{0} \models \inp$.
\end{definition}

Our definition leaves the top level formula without a set of time points. This is justified, as the only useful value there is \zeroset; it is required for satisfaction of an \ltli formula in Def.~\ref{def:ltlisemantics}.
In Remark \ref{thm:ltliproperties} we state some properties of \ltli.

\myremark%
{\thmltliproperties}%
{thm:ltliproperties}%
{Properties of \ltli}%
{\begin{inparaenum}%
\item An \ltli formula $\inpi$ s.t.~all sets of time points are $\allnats$ is equivalent to the LTL formula that one obtains from $\inpi$ by removing all sets of time points. %
\item An \ltli formula $\inpi$ with a positive (resp.~negative) polarity subformula $\prti$, where $\prti$ is neither a Boolean constant nor an atomic proposition, s.t.~all sets of time points of the top level operator of $\prti$ are $\emptyset$ is equivalent to $\inpi$ with $\prti$ replaced with $\true$ (resp.~$\false$). %
\item If $\inpi$ and $\inpqi$ are two \ltli formulas s.t.~$\inpi$ and $\inpqi$ differ only in their sets of time points, and all sets of time points in $\inpqi$ are (possibly non-strict) supersets of those in $\inpi$, then $\fbimpliesi{\inpqi}{\inpi}{\allnats}{\allnats}$. %
\item \ltli with sets of time points restricted to semilinear sets is no more expressive than \qltl (for \qltl see, e.g., \cite{EEmerson-HandbookOfTheoreticalComputerScience-1990}).%
\end{inparaenum}}
\thmltliproperties{true}

We now illustrate \ltli with an example \eqref{ex:everysecond}, \eqref{ex:everysecond:c} that is somewhat more involved than \eqref{ex:abstract}--\eqref{ex:abstract:c} in Sec.~\ref{sec:introduction}.
The example is still artificial to allow focusing on sets of time points.
\ifqaplsubmission
\else
\begin{equation}\label{ex:everysecond}
\fband{\ap}{\fband{(\fglobally{(\fbimplies{\ap}{\fnext{\fnext{\ap}}})})}{(\ffinally{(\fband{(\fbnot{\ap})}{\fnext{\fbnot{\ap}}})})}}
\end{equation}
\fi
The first conjunct, $\ap$, and the second conjunct, $\fglobally{(\fbimplies{\ap}{\fnext{\fnext{\ap}}})}$, force $\ap$ to be $\true$ at even time points.
The third conjunct, $\ffinally{(\fband{(\fbnot{\ap})}{\fnext{\fbnot{\ap}}})}$, requires that eventually $\ap$ is $\false$ at two consecutive time points.
Clearly, the first two conjuncts contradict the third, i.e., \eqref{ex:everysecond} is unsatisfiable.
We would now like to obtain small sets of time points that are still sufficient for \eqref{ex:everysecond} to be unsatisfiable.
The three conjuncts $\ap$, $\fglobally{(\fbimplies{\ap}{\fnext{\fnext{\ap}}})}$, and $\ffinally{(\fband{(\fbnot{\ap})}{\fnext{\fbnot{\ap}}})}$ are evaluated only at time point 0.
The operand of the second conjunct, $\fbimplies{\ap}{\fnext{\fnext{\ap}}}$, needs to be evaluated only at even time points and, therefore, also both operands of the $\fbimpliesname$ operator.
Consequently, it is sufficient to evaluate $\fnext{\ap}$ at odd time points and its operand, $\ap$, at even time points $> 0$.
The last conjunct is more complicated.
The operand of the $\ffinallyname$ operator has to be evaluated at every time point; otherwise, $\ffinally{(\fband{(\fbnot{\ap})}{\fnext{\fbnot{\ap}}})}$ would evaluate to $\true$.
Now note that at each time point one of the two conjuncts of $\fband{(\fbnot{\ap})}{\fnext{\fbnot{\ap}}}$ must contradict a $\ap$ induced by $\fband{\ap}{(\fglobally{(\fbimplies{\ap}{\fnext{\fnext{\ap}}})})}$.
At time point 0 this can only be the first conjunct, $\fbnot{\ap}$.
Hence, if the first conjunct, $\fbnot{\ap}$, is evaluated at even time points and the second conjunct, $\fnext{\fbnot{\ap}}$, is evaluated at odd time points, then unsatisfiability is preserved.
The resulting \ltli formula is shown in \eqref{ex:everysecond:c}.
We call \eqref{ex:everysecond:c} a \uc of \eqref{ex:everysecond} in LTL with sets of time points.
\ifqaplsubmission

{\scriptsize
\vspace{-1ex}
\hspace*{-0.0575\linewidth}
\begin{minipage}{0.34\linewidth}
\begin{equation}\label{ex:everysecond}
\fband{\ap}{\fband{(\fglobally{(\fbimplies{\ap}{\fnext{\fnext{\ap}}})})}{(\ffinally{(\fband{(\fbnot{\ap})}{\fnext{\fbnot{\ap}}})})}}
\end{equation}
\end{minipage}
\hspace{0.02\linewidth}
\begin{minipage}{0.64\linewidth}
\begin{equation}\label{ex:everysecond:c}
\fbandi{\ap}{(\fbandi{(\fgloballyi{(\fbimpliesi{\ap}{\fnexti{\fnexti{\ap}{\fslssone{2}{2}}}{\fslssone{2}{1}}}{\fslssoneone{2}}{\fslssoneone{2}})}{\fslssoneone{2}})}{(\ffinallyi{(\fbandi{(\fbnoti{\ap}{\fslssoneone{2}})}{\fnexti{\fbnoti{\ap}{\fslssone{2}{2}}}{\fslssone{2}{2}}}{\fslssoneone{2}}{\fslssone{2}{1}})}{\allnats})}{\zeroset}{\zeroset})}{\zeroset}{\zeroset}
\end{equation}
\end{minipage}
}
\else
\begin{equation}\label{ex:everysecond:c}
\fbandi{\ap}{(\fbandi{(\fgloballyi{(\fbimpliesi{\ap}{\fnexti{\fnexti{\ap}{\fslssone{2}{2}}}{\fslssone{2}{1}}}{\fslssoneone{2}}{\fslssoneone{2}})}{\fslssoneone{2}})}{(\ffinallyi{(\fbandi{(\fbnoti{\ap}{\fslssoneone{2}})}{\fnexti{\fbnoti{\ap}{\fslssone{2}{2}}}{\fslssone{2}{2}}}{\fslssoneone{2}}{\fslssone{2}{1}})}{\allnats})}{\zeroset}{\zeroset})}{\zeroset}{\zeroset}
\end{equation}
\fi

\section{\UC Extraction with Sets of Time Points}
\label{sec:coreextractioni}

In this section we show how to enhance a \uc in SNF and in LTL with the sets of time points at which its clauses or subformulas are used in its \tr proof of unsatisfiability.

\ifqaplsubmission
\else
\paragraph{\UCs in SNF with Sets of Time Points}
\fi

Let $\setclauses$ be a set of SNF clauses to which Alg.~\ref{alg:ltlsattrp} has been applied and shown unsatisfiability, let $\graph$ be the resolution graph, let $\graphp$ be the subgraph according to Def.~\ref{def:coreextraction} with corresponding \uc in SNF $\setclausesuc$, and let $\vertex_\emptyclause$ denote the vertex in the main partition that is $\vertexlabelingname$-labeled with \emptyclause.
We start in Def.~\ref{def:edgelabeling} with labeling edges of $\graphp$ with 1 if the source vertex is time-shifted one step into the future with respect to the target vertex (e.g., when a global clause with empty $\fnextname$ part is used in \refstepnx) and all other edges with 0.
Then, in Def.~\ref{def:vertexlabelingtwo}, we obtain a set of time points for each vertex in $\graphp$ by assigning time point 0 to $\vertex_\emptyclause$ (i.e., the contradiction is assumed to happen at time point 0). Any other vertex $\vertex$ is assigned the set of the sums of the time steps that occur on any path from $\vertex$ to $\vertex_\emptyclause$ in $\graphp$.

\mydefinition{Labeling Edges with Time Steps}\label{def:edgelabeling}
$\edgelabelingname$ is a labeling of the set of edges in $\graphp$, $\setedgesp$, with time steps in $\{0,1\}$ that maps an edge $\edge$ to 1 if the corresponding column 9 (\emph{t.s.~1}) or 11 (\emph{t.s.~2}) in Tab.~\ref{tab:productionrules} contains a \yes and to 0 otherwise.
\end{definition}

\mydefinition{Labeling Vertices with Sets of Time Points}\label{def:vertexlabelingtwo}
Let the edges of $\graphp$ be $\edgelabelingname$-labeled.
$\vertexlabelingtwoname$ is another labeling of the set of vertices in $\graphp$, $\setverticesp$, with sets of time points in $\mkpowerset{\allnats}$ as follows.
$\vertex_\emptyclause$ is $\vertexlabelingtwoname$-labeled with $\{0\}$.
Any other vertex $\vertex$ is $\vertexlabelingtwoname$-labeled with a set of time points $\setpos$ that contains a time point $\pos$ iff there exists a path $\ppath$ in $\graphp$ from $\vertex$ to $\vertex_\emptyclause$ such that the sum of the $\edgelabelingname$-labels of $\ppath$ is $\pos$.
\end{definition}

\begin{sloppypar}
We now continue the example in Fig.~\ref{fig:exresgraphucslsex}.
Edges in the subgraph backward reachable from $\vertex_\emptyclause$ that involve a time step of 1 between source and target vertex according to cols.~9 and 11 of Tab.~\ref{tab:productionrules} are marked red, dotted.
Backward reachable edges that involve no such time step are marked blue, dashed.
In the backward reachable subgraph there are four edges that involve a time step of 1 between source and target vertex.
Two of those originate from instances of \refloopitinitn: from $\fglobally{(\fbor{(\fbnot{a})}{\fbnot{c}})}$ in row 1 to $\fglobally{(\fnext{(\fbor{(\fbnot{a})}{\fbnot{c}})})}$ in row 5 and from $\fglobally{(\fbor{(\fbnot{b})}{\fbnot{c}})}$ in row 2 to $\fglobally{(\fnext{(\fbor{(\fbnot{b})}{\fbnot{c}})})}$ in row 5.
Two others come from instances of \refloopitsub: from $\fglobally{(\fbnot{b})}$ to $\fglobally{(\fnext{(\fbor{(\fbnot{b})}{c})})}$ and from $\fglobally{(\fbnot{a})}$ to $\fglobally{(\fnext{(\fbor{(\fbnot{a})}{c})})}$, both from row 6 to row 5.
Furthermore, there are two edges from instances of \refloopconclusiontwo that would be labeled with a time step of 1, if they were backward reachable from $\vertex_\emptyclause$: from $\fglobally{(\fbnot{b})}$ (row 6) to $\fglobally{(\fbor{(\fbnot{\fapwaitfor{c}})}{\fnext{(\fbor{(\fbnot{b})}{c})}})}$ (row 7) and from $\fglobally{(\fbnot{a})}$ (row 6) to $\fglobally{(\fbor{(\fbnot{\fapwaitfor{c}})}{\fnext{(\fbor{(\fbnot{a})}{c})}})}$ (row 7).
Figure \ref{fig:exresgraphucslsex} contains no edges induced by an instance of \refstepnx.
Notice how in each case the literals that are taken from the source vertex and put into the target vertex are in the $\fnextname$ part of the target vertex while they are not in the $\fnextname$ part of the source vertex; this is not the case for pairs of source and target vertex connected by an edge that is (or would be) labeled with time step 0.
\end{sloppypar}

Each clause $\clause$ in the backward reachable subgraph is labeled with a set of time points (shown in a black box) obtained by counting the number of red, dotted edges that are traversed on any --- possibly looping --- path from $\vertex_\clause$ to $\vertex_\emptyclause$ according to Def.~\ref{def:vertexlabelingtwo}.
For example, $a$ in row 1 can only reach $\emptyclause$ directly via a blue, dashed edge, leading to set of time points $\zeroset$ (which is the only one making sense for an initial clause; see Lemma \ref{thm:initialclauseslabeledzeroset}).
Similarly, $\fglobally{(\fbnot{a})}$ (row 8), $\fglobally{(\fbor{(\fbnot{a})}{c})}$ (row 7), and $\fglobally{(\ffinally{c})}$ (row 1) can only reach $\emptyclause$ via sequences of blue, dashed edges, so they are also labeled with $\zeroset$.
Only one of the clauses comprising the second loop search iteration (rows 5 and 6 in the light green shaded rectangle) can reach $\emptyclause$ without passing through any other clause in rows 5 or 6, namely $\fglobally{(\fbnot{a})}$ (row 6) via a sequence of blue, dashed edges.
I.e., its set of time points must contain $\zeroset$.
However, $\fglobally{(\fbnot{a})}$ is also part of the loop \ifqaplsubmission$\fglobally{(\fbnot{a})}$--$\fglobally{(\fnext{(\fbor{(\fbnot{a})}{c})})}$--$\fglobally{(\fnext{\fbnot{a}})}$--$\fglobally{(\fbnot{b})}$--$\fglobally{(\fnext{(\fbor{(\fbnot{b})}{c})})}$--$\fglobally{(\fnext{\fbnot{b}})}$--$\fglobally{(\fbnot{a})}$\else$\fglobally{(\fbnot{a})}$---$\fglobally{(\fnext{(\fbor{(\fbnot{a})}{c})})}$---$\fglobally{(\fnext{\fbnot{a}})}$---$\fglobally{(\fbnot{b})}$---$\fglobally{(\fnext{(\fbor{(\fbnot{b})}{c})})}$---$\fglobally{(\fnext{\fbnot{b}})}$---$\fglobally{(\fbnot{a})}$\fi\xspace that involves a time step of 1 between $\fglobally{(\fbnot{a})}$ and $\fglobally{(\fnext{(\fbor{(\fbnot{a})}{c})})}$ as well as between $\fglobally{(\fbnot{b})}$ and $\fglobally{(\fnext{(\fbor{(\fbnot{b})}{c})})}$.
Hence, for each even $\pos$ there exists a path such that $\fglobally{(\fbnot{a})}$ can reach $\emptyclause$ on that path and that path contains $\pos$ edges involving time steps of 1.
Consequently, $\fglobally{(\fbnot{a})}$ is labeled with $\evenset$.
The same holds for all vertices in rows 5 and 6 that are either on the loop between $\fglobally{(\fnext{(\fbor{(\fbnot{b})}{c})})}$ and $\fglobally{(\fbnot{a})}$ or backward reachable from those via blue, dashed edges: $\fglobally{(\fnext{(\fbor{(\fbnot{b})}{c})})}$, $\fglobally{(\fnext{\fbnot{b}})}$, $\fglobally{(\fnext{(\fbor{(\fbnot{b})}{\fbnot{c}})})}$, and $\fglobally{(\fbor{(\fbnot{a})}{\fnext{b}})}$.
Analogously all vertices in rows 5 and 6 that are on the loop between $\fglobally{(\fnext{(\fbor{(\fbnot{a})}{c})})}$ and  $\fglobally{(\fbnot{b})}$ or backward reachable from those via blue, dashed edges are labeled with $\oddset$: $\fglobally{(\fnext{(\fbor{(\fbnot{a})}{c})})}$, $\fglobally{(\fnext{\fbnot{a}})}$, $\fglobally{(\fbnot{b})}$, $\fglobally{(\fnext{(\fbor{(\fbnot{a})}{\fbnot{c}})})}$, and $\fglobally{(\fbor{(\fbnot{b})}{\fnext{a}})}$.
Finally, consider $\fglobally{(\fbor{(\fbnot{a})}{\fbnot{c}})}$ in row 1.
It reaches $\emptyclause$ via $\fglobally{(\fbnot{a})}$ traversing no red, dotted edge, giving $\zeroset$.
However, there is also the set of paths through the partition of the second loop search iteration, which uses $\fslssone{2}{2}$ red, dotted edges.
Taking both contributions together we obtain $\evenset$ for this clause.

From now on we assume in this section that the edges and vertices of $\graphp$ are labeled according to Def.~\ref{def:edgelabeling} and \ref{def:vertexlabelingtwo}.
The following two lemmas are needed to prove correctness of \uc extraction in SNF with sets of time points in Thm.~\ref{thm:coreiisunsat}. They can easily be proved from Def.~\ref{def:edgelabeling}, \ref{def:vertexlabelingtwo}.
Proposition \ref{thm:setsoftimepointsaresemilinearsets} establishes that the sets of time points obtained in Def.~\ref{def:vertexlabelingtwo} are semilinear (as suggested for tableaux in \cite{VSchuppan-SCP-2012}). The construction in its proof will later be a fundamental step to actually compute the sets of time points.

\mylemma%
{\thminitialclauseslabeledzeroset}%
{thm:initialclauseslabeledzeroset}%
{Sets of Time Points for Vertices Labeled with Initial Clauses are $\zeroset$}%
{Any vertex $\vertex$ in $\graphp$ that is $\vertexlabelingname$-labeled with an initial clause is $\vertexlabelingtwoname$-labeled with $\zeroset$.}%
\thminitialclauseslabeledzeroset{true}

\mylemma%
{\thmtargetclauselabeledsubsetofsourceclause}%
{thm:targetclauselabeledsubsetofsourceclause}%
{Labeling of Target Vertex is (Possibly Time-Shifted) Subset of Labeling of Source Vertex}%
{For each pair of vertices $\vertex, \vertexp$ in $\graphp$ such that there is an edge from $\vertex$ to $\vertexp$ in $\graphp$, the labeling $\fvertexlabelingtwo{\vertexp}$ is a (premise 1 of \refstepnx, \refloopitinitn, \refloopitsub, \refloopconclusiontwo: time-shifted) subset of the labeling $\fvertexlabelingtwo{\vertex}$.}
\thmtargetclauselabeledsubsetofsourceclause{true}

\myproposition%
{\thmsetsoftimepointsaresemilinearsets}%
{thm:setsoftimepointsaresemilinearsets}%
{Sets of Time Points are Semilinear Sets}%
{For each vertex $\vertex$ in $\graphp$ the labeling $\fvertexlabelingtwo{\vertex}$ is a semilinear set.}
\thmsetsoftimepointsaresemilinearsets{true}

\begin{proof}
For each vertex $\vertex$ turn the graph $\graphp$ into a transition-labeled nondeterministic finite automaton (NFA) on finite words over $\{0,1\}$ as follows:
\begin{inparaenum}[(i)]
\item The set of states is the set of vertices of the graph $\graphp$, $\setverticesp$.
\item The set of transitions is the set of \emph{reversed} edges of the graph $\graphp$.
\item The labeling of the transitions is given by the $\edgelabelingname$-labeling of the corresponding edges.
\item The (only) initial state is $\vertex_\emptyclause$.
\item The (only) final state is $\vertex$.
\end{inparaenum}
Now it's clear from Def.~\ref{def:vertexlabelingtwo} that the $\vertexlabelingtwoname$-labeling of the vertex $\vertex$ is the Parikh image of the letter 1 of the regular language given by the automaton. The claim follows from Parikh's theorem \cite{RParikh-JACM-1966}.
\end{proof}

We now define \ucs in SNF with sets of time points.
To simplify notation we first define what it means to assign a set of time points to an SNF clause (Def.~\ref{def:snfclauseswithsetsoftimepoints}).
The definition of a \uc in SNF with sets of time points is then immediate in Def.~\ref{def:coreextractionltli}.
Given the proof of Thm.~\ref{thm:coreisunsat} (see \cite{VSchuppan-NFM-2013-full-arXiv}) the proof of correctness in Thm.~\ref{thm:coreiisunsat} (in \refformalcoreextractioni) can focus on why the construction remains correct with sets of time points.
In Prop.~\ref{thm:coreicomplexity} we state an upper bound on the complexity of extracting a \uc in SNF with sets of time points.

\mydefinition{SNF Clauses with Sets of Time Points}\label{def:snfclauseswithsetsoftimepoints}
Let $\setpos$ be a set of time points. Let $\clause$ be an SNF clause. Then \emph{$\clause$ with set of time points $\setpos$}, $\fclausei{\clause}{\setpos}$, is the following $\ltli$ formula:\footnote{In this definition $(\fbnotname)$ indicates a negation that may or may not be present.}
{\ifqaplsubmission\scriptsize\fi
\[
\fclausei{\clause}{\setpos} = \left\{
\begin{array}{l}
\ifqaplsubmission
\ficlausei{\fbori{(\fbnoti{}{\setpos})\ap_1}{\fbori{\ldots}{(\fbnoti{}{\setpos})\ap_\maxind}{\setpos}{\setpos}}{\setpos}{\setpos}} \hspace{0.5em} \mbox{if } c = \ficlause{\fbor{(\fbnotname)\ap_1}{\fbor{\ldots}{(\fbnotname)\ap_\maxind}}} \mbox{ is an initial clause; or}\\[3ex]
(\fgloballyi{(\fbori{\fbori{(\fbnoti{}{\setpos})\ap_1}{\fbori{\ldots}{(\fbnoti{}{\setpos})\ap_\maxind}{\setpos}{\setpos}}{\setpos}{\setpos}}{(\fnexti{\fbori{(\fbnoti{}{\setpos + 1})\app_1}{\fbori{\ldots}{(\fbnoti{}{\setpos + 1})\app_\maxindp}{\setpos + 1}{\setpos + 1}}{\setpos + 1}{\setpos + 1}}{\setpos + 1})}{\setpos}{\setpos}}{\setpos}))\\[2ex]
\hspace*{1em}\mbox{\begin{minipage}{0.85\linewidth}if $c = \fgnxclause{\fbor{(\fbnotname)\ap_1}{\fbor{\ldots}{(\fbnotname)\ap_\maxind}}}{\fbor{(\fbnotname)\app_1}{\fbor{\ldots}{(\fbnotname)\app_\maxindp}}}$ is a global clause; or\end{minipage}}\\[3ex]
\feclausei{\fbori{(\fbnoti{}{\setpos})\ap_1}{\fbori{\ldots}{(\fbnoti{}{\setpos})\ap_\maxind}{\setpos}{\setpos}}{\setpos}{\setpos}}{(\fbnoti{}{[\fmin{\setpos},\infty)})\apres}{\setpos}{[\fmin{\setpos},\infty)} \hspace{0.5em} \mbox{if } c = \feclause{\fbor{(\fbnotname)\ap_1}{\fbor{\ldots}{(\fbnotname)\ap_\maxind}}}{(\fbnotname)\apres} \mbox{ is an eventuality clause.}\\
\else
\ficlausei{\fbori{(\fbnoti{}{\setpos})\ap_1}{\fbori{\ldots}{(\fbnoti{}{\setpos})\ap_\maxind}{\setpos}{\setpos}}{\setpos}{\setpos}}\\[2ex]
\hspace*{1em}\mbox{\begin{minipage}{0.85\linewidth}if $c = \ficlause{\fbor{(\fbnotname)\ap_1}{\fbor{\ldots}{(\fbnotname)\ap_\maxind}}}$ is an initial clause; or\end{minipage}}\\[3ex]
(\fgloballyi{(\fbori{\fbori{(\fbnoti{}{\setpos})\ap_1}{\fbori{\ldots}{(\fbnoti{}{\setpos})\ap_\maxind}{\setpos}{\setpos}}{\setpos}{\setpos}}{(\fnexti{\fbori{(\fbnoti{}{\setpos + 1})\app_1}{\fbori{\ldots}{(\fbnoti{}{\setpos + 1})\app_\maxindp}{\setpos + 1}{\setpos + 1}}{\setpos + 1}{\setpos + 1}}{\setpos + 1})}{\setpos}{\setpos}}{\setpos}))\\[2ex]
\hspace*{1em}\mbox{\begin{minipage}{0.85\linewidth}if $c = \fgnxclause{\fbor{(\fbnotname)\ap_1}{\fbor{\ldots}{(\fbnotname)\ap_\maxind}}}{\fbor{(\fbnotname)\app_1}{\fbor{\ldots}{(\fbnotname)\app_\maxindp}}}$ is a global clause; or\end{minipage}}\\[3ex]
\feclausei{\fbori{(\fbnoti{}{\setpos})\ap_1}{\fbori{\ldots}{(\fbnoti{}{\setpos})\ap_\maxind}{\setpos}{\setpos}}{\setpos}{\setpos}}{(\fbnoti{}{[\fmin{\setpos},\infty)})\apres}{\setpos}{[\fmin{\setpos},\infty)}\\[2ex]
\hspace*{1em}\mbox{\begin{minipage}{0.85\linewidth}if $c = \feclause{\fbor{(\fbnotname)\ap_1}{\fbor{\ldots}{(\fbnotname)\ap_\maxind}}}{(\fbnotname)\apres}$ is an eventuality clause.\end{minipage}}\\
\fi
\end{array}
\right.
\]
}
\end{definition}

\mydefinition{\UC in SNF with Sets of Time Points}\label{def:coreextractionltli}
Let $\clause_{1,1}, \ldots, \clause_{1,\maxind_1}$ be the initial clauses in $\setclausesuc$, \ifqaplsubmission$\clause_{2,1}, \ldots, \clause_{2,\maxind_2}$\else$\clause_{2,1}$, $\ldots$, $\clause_{2,\maxind_2}$\fi\xspace the global clauses in $\setclausesuc$, and $\clause_{3,1}$, $\ldots$, $\clause_{3,\maxind_3}$ the eventuality clauses in $\setclausesuc$.
Let $\vertex_{\nat,\natp}$ be the unique vertex in the main partition $\mainpartition$ of $\graphp$ $\vertexlabelingname$-labeled with clause $\clause_{\nat,\natp}$.
Let $\setpos_{\nat,\natp}$ be the set of time points that vertex $\vertex_{\nat,\natp}$ is $\vertexlabelingtwoname$-labeled with in $\graphp$.
The \emph{\uc of $\setclauses$ in SNF with sets of time points}, $\inpuci$, is given by
{\ifqaplsubmission\scriptsize\fi
\def\negspc{\!\!\!}
\[
\begin{array}{c}
  \fbandi{\fclausei{c_{1,1}}{\setpos_{1,1}}}{\fbandi{\negspc\ldots\negspc}{\fclausei{c_{1,\maxind_1}}{\setpos_{1,\maxind_1}}}{\zeroset}{\zeroset}}{\zeroset}{\zeroset}
  \fbandi{\hspace{0em}}{\hspace{0em}}{\zeroset}{\zeroset}
  \fbandi{\fclausei{c_{2,1}}{\setpos_{2,1}}}{\fbandi{\negspc\ldots\negspc}{\fclausei{c_{2,\maxind_2}}{\setpos_{2,\maxind_2}}}{\zeroset}{\zeroset}}{\zeroset}{\zeroset}
  \fbandi{\hspace{0em}}{\hspace{0em}}{\zeroset}{\zeroset}
  \fbandi{\fclausei{c_{3,1}}{\setpos_{3,1}}}{\fbandi{\negspc\ldots\negspc}{\fclausei{c_{3,\maxind_3}}{\setpos_{3,\maxind_3}}}{\zeroset}{\zeroset}}{\zeroset}{\zeroset}.
\end{array}
\]
}
\end{definition}

\mytheorem%
{\thmcoreiisunsat}%
{thm:coreiisunsat}%
{Unsatisfiability of \UC in SNF with Sets of Time Points}
{Let $\inpuci$ be the \uc of $\setclauses$ in SNF with sets of time points. Then $\inpuci$ is unsatisfiable.}
\thmcoreiisunsat{true}

\myproposition
{\thmcoreicomplexity}%
{thm:coreicomplexity}%
{Complexity of \UC Extraction with Sets of Time Points}%
{Let $\inpuci$ be the \uc of $\setclauses$ in SNF with sets of time points. Construction of $\inpuci$ from $\graphp$ can be performed in time $\fbigo{\fcardinality{\setverticesp}^3 + \fcardinality{\setverticesp}^2 \cdot \fcardinality{\setclauses}}$.}
\thmcoreicomplexity{true}

\begin{proof}
(Sketch)
Construct an NFA from $\graphp$ along the lines of the proof of Prop.~\ref{thm:setsoftimepointsaresemilinearsets}.
Turn the NFA into a unary NFA by regarding edges $\edgelabelingname$-labeled with 0 as $\epsilon$-edges and making the NFA $\epsilon$-free (e.g., \cite{JHopcroftJUllman-1979}).
Finally, use an algorithm by Gawrychowski \cite{PGawrychowski-CIAA-2011} extended to handle all final states in parallel to compute sets of time points.
\end{proof}

We now apply Def.~\ref{def:coreextractionltli} to the example in Fig.~\ref{fig:exresgraphucslsex} and obtain \eqref{ex:exresgraphucslsex:csnf} as a \uc in SNF with sets of time points.
Notice, that all occurrences of $a$ occur at even time points and how both occurrences of $b$ interact at odd time points.
Moreover, the last clause shows that only a single occurrence of $c$ is required for unsatisfiability.
Finally, the fourth clause has $\fbnot{c}$ at even time points, while the fifth clause becomes relevant at odd time points; thus all potential occurrences of $c$ are covered.
This concludes this example.
\ifqaplsubmission

{\scriptsize
\begin{equation}\label{ex:exresgraphucslsex:csnf}
\begin{array}{ll}
&
\fbandi{a}{\fbandi{(\fgloballyi{(\fbori{(\fbnoti{a}{\evenset})}{\fnexti{b}{\oddset}}{\evenset}{\evenset})}{\evenset})}{(\fgloballyi{(\fbori{(\fbnoti{b}{\oddset})}{\fnexti{a}{\fslssone{2}{2}}}{\oddset}{\oddset})}{\oddset})}{\zeroset}{\zeroset}}{\zeroset}{\zeroset}
\\[2ex]
\fbandi{\hspace*{0em}}{\hspace*{0em}}{\zeroset}{\zeroset}
&
\fbandi{(\fgloballyi{(\fbori{(\fbnoti{a}{\evenset})}{\fbnoti{c}{\evenset}}{\evenset}{\evenset})}{\evenset})}{\fbandi{(\fgloballyi{(\fbori{(\fbnoti{c}{\oddset})}{\fnexti{\fbnoti{a}{\fslssone{2}{2}}}{\fslssone{2}{2}}}{\oddset}{\oddset})}{\oddset})}{(\fgloballyi{(\ffinallyi{c}{\allnats})}{\zeroset})}{\zeroset}{\zeroset}}{\zeroset}{\zeroset}
\end{array}
\end{equation}
}
\\[-0.5ex]
\else
\begin{equation}\label{ex:exresgraphucslsex:csnf}
\begin{array}{c}
\fbandi{\fbandi{a}{\fbandi{(\fgloballyi{(\fbori{(\fbnoti{a}{\evenset})}{\fnexti{b}{\oddset}}{\evenset}{\evenset})}{\evenset})}{(\fgloballyi{(\fbori{(\fbnoti{b}{\oddset})}{\fnexti{a}{\fslssone{2}{2}}}{\oddset}{\oddset})}{\oddset})}{\zeroset}{\zeroset}}{\zeroset}{\zeroset}}{\hspace*{0em}}{\zeroset}{\zeroset}
\\
\fbandi{(\fgloballyi{(\fbori{(\fbnoti{a}{\evenset})}{\fbnoti{c}{\evenset}}{\evenset}{\evenset})}{\evenset})}{\fbandi{(\fgloballyi{(\fbori{(\fbnoti{c}{\oddset})}{\fnexti{\fbnoti{a}{\fslssone{2}{2}}}{\fslssone{2}{2}}}{\oddset}{\oddset})}{\oddset})}{(\fgloballyi{(\ffinallyi{c}{\allnats})}{\zeroset})}{\zeroset}{\zeroset}}{\zeroset}{\zeroset}
\end{array}
\end{equation}
\fi

\ifqaplsubmission
\else
\paragraph{\UCs in LTL with Sets of Time Points}
\fi

Definition \ref{def:ltlicoreextraction} adds sets of time points to a \uc in LTL by transferring them from a \uc in SNF with time points to a \uc in LTL. The proof idea for Thm.~\ref{thm:ltlicoreisunsat} (in \refformalcoreextractioni) is similar to that of Thm.~\ref{thm:ltlcoreisunsat} (see \cite{VSchuppan-NFM-2013-full-arXiv}), but in addition we need to define a translation from the corresponding fragment of \ltli to SNF with sets of time points, which must be shown to be satisfiability- but not unsatisfiability-preserving.

\mydefinition{Mapping a \UC in SNF with Sets of Time Points to a \UC in LTL with Sets of Time Points} \label{def:ltlicoreextraction}
Let $\inp$ be an unsatisfiable LTL formula, let $\fdCNF{\inp}$ be its SNF, let $\inpuc$ be the \uc of $\inp$ in LTL, and let $\inpuci$ be the \uc of $\fdCNF{\inp}$ in SNF with sets of time points. Construct the \emph{\uc of $\inp$ in LTL with sets of time points}, $\inppuci$, by assigning a set of time points $\setpos$ to each occurrence of a subformula $\prt$ in $\inpuc$ as follows. Let $\setposp, \setpospp, \ldots$ be the sets of time points of the occurrences of the proposition $\fdCNFvar{\prt}$ in $\inpuci$ that are marked {\color{blue}\setlength\fboxsep{1pt}\fbox{blue boxed}} in Tab.~\ref{tab:ltltosnf}. Then assign the occurrence of $\prt$ in $\inpuc$ the set of time points $\setpos$ that is the union of $\setposp, \setpospp, \ldots$.
\end{definition}

\mytheorem%
{\thmltlicoreisunsat}%
{thm:ltlicoreisunsat}%
{Unsatisfiability of \UC in LTL with Sets of Time Points}%
{Let $\inp$ be an unsatisfiable LTL formula, and let $\inppuci$ be the \uc of $\inp$ in LTL with sets of time points. Then $\inppuci$ is unsatisfiable.}
\thmltlicoreisunsat{true}

It's easy to see that no subformula in \eqref{ex:abstract} or \eqref{ex:everysecond} can be replaced with $\true$ (for positive polarity occurrences) or $\false$ (for negative polarity occurrences) without making \eqref{ex:abstract} or \eqref{ex:everysecond} satisfiable. I.e., \eqref{ex:abstract} or \eqref{ex:everysecond} are the only \ucs of themselves according to Def.~10 in \cite{VSchuppan-SCP-2012} (and, hence, according to Def.~\ref{def:ltlcoreextraction}). The corresponding \ucs in LTL with sets of time points in \eqref{ex:abstract:c} and \eqref{ex:everysecond:c} show that \ucs with sets of time points can be more fine-grained than \ucs without.

\section{Example}\label{sec:example}

In this section we present an example that shows the utility of \ucs with sets of time points for debugging that is closer to a real world situation.
The \ucs in this as well as in all other examples in this paper were obtained with our implementation, possibly except for minor rewriting.

The example \eqref{ex:lift} in Fig.~\ref{fig:ex:lift} reuses the example of a lift specification from \cite{VSchuppan-NFM-2013-full-arXiv} (originally adapted from \cite{AHarding-PhDThesis-2005}) but extends it with sets of time points to show that understanding the presence of a problem becomes easier.
The lift has two floors, indicated by $f_0$ and $f_1$. On each floor there is a button to call the lift ($b_0$, $b_1$). $sb$ is $\true$ if some button is pressed.
If the lift moves up, then $up$ must be $\true$; if it moves down, then $up$ must be $\false$.
$u$ switches turns between actions by users of the lift ($u$ is $\true$) and actions by the lift ($u$ is $\false$).
For a more detailed explanation we refer to \cite{AHarding-PhDThesis-2005}.
\begin{figure}
\ifqaplsubmission
{\scriptsize
\begin{subequations}\label{ex:lift}
\hspace*{-0.015\linewidth}
\begin{minipage}{0.56\linewidth}
\begin{align}
&\fband{(\fbnot{u})}{\fband{(f_0)}{\fband{(\fbnot{b_0})}{\fband{(\fbnot{b_1})}{(\fbnot{up})}}}} \\[-0.75ex]
\fbandname\; &(\fglobally{(\fband{(\fbimplies{u}{\fbnot{\fnext{u}}})}{(\fbimplies{(\fbnot{\fnext{u}})}{u})})}) \\[-0.75ex]
\fbandname\; &(\fglobally{(\fbimplies{f_0}{\fbnot{f_1}})}) \\[-0.75ex]
\fbandname\; &(\fglobally{(\fband{(\fbimplies{f_0}{\fnext{(\fbor{f_0}{f_1})}})}{(\fbimplies{f_1}{\fnext{(\fbor{f_0}{f_1})}})})}) \\[-0.75ex]
\fbandname\; &(\fglobally{(\fbimplies{u}{(\fband{\fband{(\fbimplies{f_0}{\fnext{f_0}})}{(\fbimplies{(\fnext{f_0})}{f_0})}}{\fband{(\fbimplies{f_1}{\fnext{f_1}})}{(\fbimplies{(\fnext{f_1})}{f_1})}})})}) \\[-0.75ex]
\fbandname\; &
\begin{minipage}{0.88\linewidth}
(\fglobally{({(\fbimplies{(\fbnot{u})}{\hspace{0em}}}} \\
\hspace*{2em}(\fband{\fband{(\fbimplies{b_0}{\fnext{b_0}})}{(\fbimplies{(\fnext{b_0})}{b_0})}}{\fband{(\fbimplies{b_1}{\fnext{b_1}})}{(\fbimplies{(\fnext{b_1})}{b_1})}}))))
\end{minipage} \\[-0.75ex]
\fbandname\; &(\fglobally{(\fband{(\fbimplies{(\fband{b_0}{\fbnot{f_0}})}{\fnext{b_0}})}{(\fbimplies{(\fband{b_1}{\fbnot{f_1}})}{\fnext{b_1}})})})
\end{align}
\end{minipage}
\hspace{0.0025\linewidth}
\begin{minipage}{0.43\linewidth}
\begin{align}
\fbandname\; &(\fglobally{(\fbimplies{(\fband{f_0}{\fnext{f_0}})}{(\fband{(\fbimplies{up}{\fnext{up}})}{(\fbimplies{(\fnext{up})}{up})})})}) \\[-0.75ex]
\fbandname\; &(\fglobally{(\fbimplies{(\fband{f_1}{\fnext{f_1}})}{(\fband{(\fbimplies{up}{\fnext{up}})}{(\fbimplies{(\fnext{up})}{up})})})}) \\[-0.75ex]
\fbandname\; &(\fglobally{(\fband{(\fbimplies{(\fband{f_0}{\fnext{f_1}})}{up})}{(\fbimplies{(\fband{f_1}{\fnext{f_0}})}{\fbnot{up}})})}) \\[-0.75ex]
\fbandname\; &(\fglobally{(\fband{(\fbimplies{sb}{(\fbor{b_0}{b_1})})}{(\fbimplies{(\fbor{b_0}{b_1})}{sb})})}) \\[-0.75ex]
\fbandname\; &(\fglobally{((\fbimplies{(\fband{f_0}{\fbnot{sb}})}{(\funtil{f_0}{(\freleases{sb}{(\fband{(\ffinally{f_0})}{(\fbnot{up})})})})}))} \\[-0.75ex]
\fbandname\; &(\fglobally{((\fbimplies{(\fband{f_1}{\fbnot{sb}})}{(\funtil{f_1}{(\freleases{sb}{(\fband{(\ffinally{f_0})}{(\fbnot{up})})})})}))} \\[-0.75ex]
\fbandname\; &(\fglobally{(\fband{(\fbimplies{b_0}{\ffinally{f_0}})}{(\fbimplies{b_1}{\ffinally{f_1}})})})
\end{align}
\end{minipage}
\end{subequations}
}
\else
\begin{subequations}\label{ex:lift}
\begin{align}
&\fband{(\fbnot{u})}{\fband{(f_0)}{\fband{(\fbnot{b_0})}{\fband{(\fbnot{b_1})}{(\fbnot{up})}}}} \\[-0.75ex]
\fbandname\; &(\fglobally{(\fband{(\fbimplies{u}{\fbnot{\fnext{u}}})}{(\fbimplies{(\fbnot{\fnext{u}})}{u})})}) \\[-0.75ex]
\fbandname\; &(\fglobally{(\fbimplies{f_0}{\fbnot{f_1}})}) \\[-0.75ex]
\fbandname\; &(\fglobally{(\fband{(\fbimplies{f_0}{\fnext{(\fbor{f_0}{f_1})}})}{(\fbimplies{f_1}{\fnext{(\fbor{f_0}{f_1})}})})}) \\[-0.75ex]
\fbandname\; &(\fglobally{(\fbimplies{u}{(\fband{\fband{(\fbimplies{f_0}{\fnext{f_0}})}{(\fbimplies{(\fnext{f_0})}{f_0})}}{\fband{(\fbimplies{f_1}{\fnext{f_1}})}{(\fbimplies{(\fnext{f_1})}{f_1})}})})}) \\[-0.75ex]
\fbandname\; &(\fglobally{({(\fbimplies{(\fbnot{u})}{(\fband{\fband{(\fbimplies{b_0}{\fnext{b_0}})}{(\fbimplies{(\fnext{b_0})}{b_0})}}{\fband{(\fbimplies{b_1}{\fnext{b_1}})}{(\fbimplies{(\fnext{b_1})}{b_1})}})})})}) \\[-0.75ex]
\fbandname\; &(\fglobally{(\fband{(\fbimplies{(\fband{b_0}{\fbnot{f_0}})}{\fnext{b_0}})}{(\fbimplies{(\fband{b_1}{\fbnot{f_1}})}{\fnext{b_1}})})}) \\[-0.75ex]
\fbandname\; &(\fglobally{(\fbimplies{(\fband{f_0}{\fnext{f_0}})}{(\fband{(\fbimplies{up}{\fnext{up}})}{(\fbimplies{(\fnext{up})}{up})})})}) \\[-0.75ex]
\fbandname\; &(\fglobally{(\fbimplies{(\fband{f_1}{\fnext{f_1}})}{(\fband{(\fbimplies{up}{\fnext{up}})}{(\fbimplies{(\fnext{up})}{up})})})}) \\[-0.75ex]
\fbandname\; &(\fglobally{(\fband{(\fbimplies{(\fband{f_0}{\fnext{f_1}})}{up})}{(\fbimplies{(\fband{f_1}{\fnext{f_0}})}{\fbnot{up}})})}) \\[-0.75ex]
\fbandname\; &(\fglobally{(\fband{(\fbimplies{sb}{(\fbor{b_0}{b_1})})}{(\fbimplies{(\fbor{b_0}{b_1})}{sb})})}) \\[-0.75ex]
\fbandname\; &(\fglobally{((\fbimplies{(\fband{f_0}{\fbnot{sb}})}{(\funtil{f_0}{(\freleases{sb}{(\fband{(\ffinally{f_0})}{(\fbnot{up})})})})}))} \\[-0.75ex]
\fbandname\; &(\fglobally{((\fbimplies{(\fband{f_1}{\fbnot{sb}})}{(\funtil{f_1}{(\freleases{sb}{(\fband{(\ffinally{f_0})}{(\fbnot{up})})})})}))} \\[-0.75ex]
\fbandname\; &(\fglobally{(\fband{(\fbimplies{b_0}{\ffinally{f_0}})}{(\fbimplies{b_1}{\ffinally{f_1}})})})
\end{align}
\end{subequations}
\fi
\caption{\label{fig:ex:lift} A lift specification.}
\end{figure}

We first assume that an engineer is interested in seeing whether it is possible that $b_1$ is continuously pressed \eqref{ex:lift:s1}. As the \uc \eqref{ex:lift:c1} shows this is impossible as $b_1$ must be $\false$ at time point 0. Notice that \eqref{ex:lift:c1} indicates that the argument of the $\fgloballyname$ operator is only needed at time point 0 (trivial to see in this case).
\ifqaplsubmission
\\[0.5ex]
{\scriptsize
\begin{minipage}{0.3\linewidth}
\begin{equation}\label{ex:lift:s1}
\fglobally{b_1}
\end{equation}
\end{minipage}
\hspace{0.2\linewidth}
\begin{minipage}{0.4\linewidth}
\begin{equation}\label{ex:lift:c1}
\fbandi{(\fbnoti{b_1}{\zeroset})}{\fgloballyi{b_1}{\zeroset}}{\zeroset}{\zeroset}
\end{equation}
\end{minipage}
}
\\[-0.5ex]
\else
\begin{gather}
\fglobally{b_1}\label{ex:lift:s1}
\\
\fbandi{(\fbnoti{b_1}{\zeroset})}{\fgloballyi{b_1}{\zeroset}}{\zeroset}{\zeroset}\label{ex:lift:c1}
\end{gather}
\fi

Now the engineer modifies her query such that $b_1$ is pressed only from time point 1 on \eqref{ex:lift:s2}. That is impossible, too; as the \uc in \eqref{ex:lift:c2} shows also this time the press of $b_1$ is required only at one time point.
\ifqaplsubmission
\\[0.5ex]
{\scriptsize
\begin{minipage}{0.15\linewidth}
\begin{equation}\label{ex:lift:s2}
\fnext{\fglobally{b_1}}
\end{equation}
\end{minipage}
\hspace{0.08\linewidth}
\begin{minipage}{0.75\linewidth}
\begin{equation}\label{ex:lift:c2}
\fbandi{(\!\fbnoti{\!u}{\zeroset})\!}{\!(\fbandi{(\!\fbnoti{\!b_1}{\zeroset})\!}{\!(\fbandi{(\!\fgloballyi{(\fbimpliesi{(\fbnoti{\!u}{\zeroset})\!}{\!(\fbimpliesi{(\!\fnexti{\!b_1}{\fslssonetwo{1}})\!}{\!b_1}{\zeroset}{\zeroset})}{\zeroset}{\zeroset})}{\zeroset})\!}{\!(\!\fnexti{\!\fgloballyi{\!b_1}{\fslssonetwo{1}}}{\fslssonetwo{1}})}{\zeroset}{\zeroset})}{\zeroset}{\zeroset})}{\zeroset}{\zeroset}
\end{equation}
\end{minipage}
}
\\[-0.5ex]
\else
\begin{gather}
\fnext{\fglobally{b_1}}\label{ex:lift:s2}
\\
\fbandi{(\!\fbnoti{\!u}{\zeroset})\!}{\!(\fbandi{(\!\fbnoti{\!b_1}{\zeroset})\!}{\!(\fbandi{(\!\fgloballyi{(\fbimpliesi{(\fbnoti{\!u}{\zeroset})\!}{\!(\fbimpliesi{(\!\fnexti{\!b_1}{\fslssonetwo{1}})\!}{\!b_1}{\zeroset}{\zeroset})}{\zeroset}{\zeroset})}{\zeroset})\!}{\!(\!\fnexti{\!\fgloballyi{\!b_1}{\fslssonetwo{1}}}{\fslssonetwo{1}})}{\zeroset}{\zeroset})}{\zeroset}{\zeroset})}{\zeroset}{\zeroset}\label{ex:lift:c2}
\end{gather}
\fi

The engineer now tries to have $b_1$ pressed only from time point 2 on and also obtains a \uc that needs $b_1$ pressed only at a single time point (not shown). She becomes suspicious and checks whether $b_1$ can be pressed at all. She now sees that $b_1$ cannot be pressed at any time point and, therefore, this specification of a lift must contain a bug. This example illustrates the benefits of \ucs with sets of time points, as \eqref{ex:lift:c1} and \eqref{ex:lift:c2} make it clear that $b_1$ being $\true$ is only needed at a single time point for unsatisfiability.

For an example showing disjuncts of an invariant holding at different time points and for an example from the business process domain see \refmoreexamples.

\section{Experimental Evaluation}
\label{sec:experimentalevaluation}

\ifqaplsubmission
\else
Our implementation, examples, and log files are available from \cite{paperwebpage}.
\fi

\ifqaplsubmission
\else
\paragraph{Implementation}
\fi

We use the version of \trp extended with extraction of \ucs from \cite{VSchuppan-NFM-2013-full-arXiv} as the basis for our implementation.
\ifqaplsubmission
\else
For data structures we used C++ Standard Library containers (e.g., \cite{AStepanovMLee-HPLaboratoriesTR-1995,NJosuttis-2012}), for graph operations the Boost Graph Library \cite{boostgraphlibrary,JSiekLLeeALumsdaine-2001}.
\fi
\ifqaplsubmission
\else

\paragraph{Algorithms for Extracting Sets of Time Points}

\fi
We implemented extraction of sets of time points along the lines of the proofs of Prop.~\ref{thm:setsoftimepointsaresemilinearsets}, \ref{thm:coreicomplexity}.
To make an NFA $\epsilon$-free we use a standard algorithm that performs DFS from each state to find the sets of states that are reachable via a sequence of $\epsilon$-edges, inserts $1$-edges between pairs of vertices $\vertex$, $\vertexp$ such that $\vertex$ can reach $\vertexp$ by reading $\epsilon^* 1 \epsilon^*$, and removes $\epsilon$-edges (e.g., \cite{JHopcroftJUllman-1979}).
To compute Parikh images for unary NFAs we implemented an algorithm by Gawrychowski \cite{PGawrychowski-CIAA-2011} and one by Sawa \cite{ZSawa-FundamentaInformaticae-2013}.
Both assume a single set of final states leading to a single Parikh image. We, however, have one final state for each SNF clause in the \uc in SNF, each of which we need to assign a separate Parikh image.
We adapted Gawrychowski's algorithm to our setting by computing the Parikh images for different final states in a single run of the algorithm.
Similarly, we optimized Sawa's algorithm by computing parts that are common for different final states only once and by heuristically accelerating some of its steps.

\ifqaplsubmission
\else
\paragraph{Benchmarks}
\fi

Our examples are based on \cite{VSchuppanLDarmawan-ATVA-2011}.
In categories \benchmark{crafted} and \benchmark{random} and in family \benchmark{forobots} we considered all unsatisfiable instances from \cite{VSchuppanLDarmawan-ATVA-2011}.
The version of \benchmark{alaska\_lift} used here contains a small bug fix: in \cite{MDeWulfLDoyenNMaquetJRaskin-TACAS-2008,VSchuppanLDarmawan-ATVA-2011} the subformula $\fnext{u}$ was erroneously written as literal $Xu$.
Combining 2 variants of \benchmark{alaska\_lift} with 3 different scenarios we obtain 6 subfamilies of \benchmark{alaska\_lift}.
For \benchmark{anzu\_genbuf} we invented 3 scenarios to obtain 3 subfamilies.
For all benchmark families that consist of a sequence of instances of increasing difficulty we stopped after two instances that could not be solved due to time or memory out.
Some instances were simplified to $\false$ during the translation from LTL to SNF; these instances were discarded.
In Tab.~\ref{tab:benchmarks} we give an overview of the benchmark families.
Columns 1--3 give the category, name, and the source of the family.
Columns 4--6 list the numbers of instances that were solved by our implementation with \uc extraction without sets of time points, with \uc extraction with sets of time points using Gawrychowski's algorithm, and  with \uc extraction with sets of time points using Sawa's algorithm.
Column 7 indicates the size (number of nodes in the syntax tree) of the largest instance solved with \uc extraction without sets of time points.

\begin{table}[t]
\centering
{\tiny
\ifqaplsubmission
\begin{tabular}{||@{\hspace{0.75em}}l@{\hspace{0.75em}}|@{\hspace{0.75em}}l@{\hspace{0.75em}}|@{\hspace{0.75em}}l@{\hspace{0.75em}}|@{\hspace{0.75em}}r@{\hspace{0.75em}}|@{\hspace{0.75em}}r@{\hspace{0.75em}}|@{\hspace{0.75em}}r@{\hspace{0.75em}}|@{\hspace{0.75em}}r@{\hspace{0.75em}}||}
\else
\begin{tabular}{||l|l|l|r|r|r|r||}
\fi
\hline
\hline
\ifqaplsubmission
category &
family &
source &
\# solved \uc w/o s.o.t.p. &
\# solved \uc w/ s.o.t.p.~(Gawrychowski) &
\# solved \uc w/ s.o.t.p.~(Sawa) &
$|\mbox{largest solved}|$
\\
\else
category &
family &
source &
\multicolumn{3}{|c|}{\# solved} &
$|\mbox{largest solved}|$
\\
&
&
&
\uc w/o s.o.t.p. &
\uc w/ s.o.t.p.~(Gawrychowski) &
\uc w/ s.o.t.p.~(Sawa) &
\\
\fi
\hline
&
\benchmark{alaska\_lift} &
\cite{AHarding-PhDThesis-2005,MDeWulfLDoyenNMaquetJRaskin-TACAS-2008} &
72 &
73 &
73 &
4605
\\
\benchmark{application} &
\benchmark{anzu\_genbuf} &
\cite{RBloemSGallerBJobstmannNPitermanAPnueliMWeiglhofer-COCV-2007} &
16 &
16 &
16 &
1924
\\
&
\benchmark{forobots} &
\cite{ABehdennaCDixonMFisher-IntelligentComputingAndCybernetics-2009} &
25 &
25 &
25 &
635
\\
\hline
&
\benchmark{schuppan\_O1formula} &
\cite{VSchuppanLDarmawan-ATVA-2011} &
27 &
27 &
27 &
4006
\\
\benchmark{crafted} &
\benchmark{schuppan\_O2formula} &
\cite{VSchuppanLDarmawan-ATVA-2011} &
8 &
7 &
7 &
91
\\
&
\benchmark{schuppan\_phltl} &
\cite{VSchuppanLDarmawan-ATVA-2011} &
4 &
4 &
4 &
125
\\
\hline
&
\benchmark{rozier\_formulas} &
\cite{KRozierMVardi-STTT-2010} &
62 &
62 &
62 &
157
\\
\raisebox{1.5ex}[-1.5ex]{\benchmark{random}} &
\benchmark{trp} &
\cite{UHustadtRSchmidt-KR-2002} &
397 &
397 &
397 &
1421 \\
\hline
\hline
\end{tabular}
}
\caption{\label{tab:benchmarks}Overview of benchmark families.}
\end{table}

\ifqaplsubmission
\else
\paragraph{Setup}
\fi

The experiments were performed on a laptop with Intel Core i7 M 620 processor at 2 GHz running Ubuntu 12.04. Run time and memory usage were measured with \tool{run} \cite{run}. The time and memory limits were 600 seconds and 6 GB.

\ifqaplsubmission
\else
\paragraph{Results}
\fi

In Fig.~\ref{fig:overheadall} (a) and (b) we show the overhead that is incurred by extracting \ucs with sets of time points.
Figure~\ref{fig:overheadall} (c) and (d) compare using Gawrychowski's and Sawa's algorithm for computing sets of time points.
In Tab.~\ref{tab:slsdistribution} we show which sets of time points occur in the \ucs of which benchmark families.

{
\newcommand{\tabfield}[1]{%
\begin{minipage}{0.225\linewidth}%
\begin{center}%
\includegraphics[type=pdf,ext=.pdf,read=.pdf,scale=0.375,trim=1.5cm 0.0cm 1.6cm 0.3cm,clip]{#1}%
\end{center}%
\end{minipage}%
}%

\begin{figure}[t]
\begin{tabular}{cccc}
\tabfield{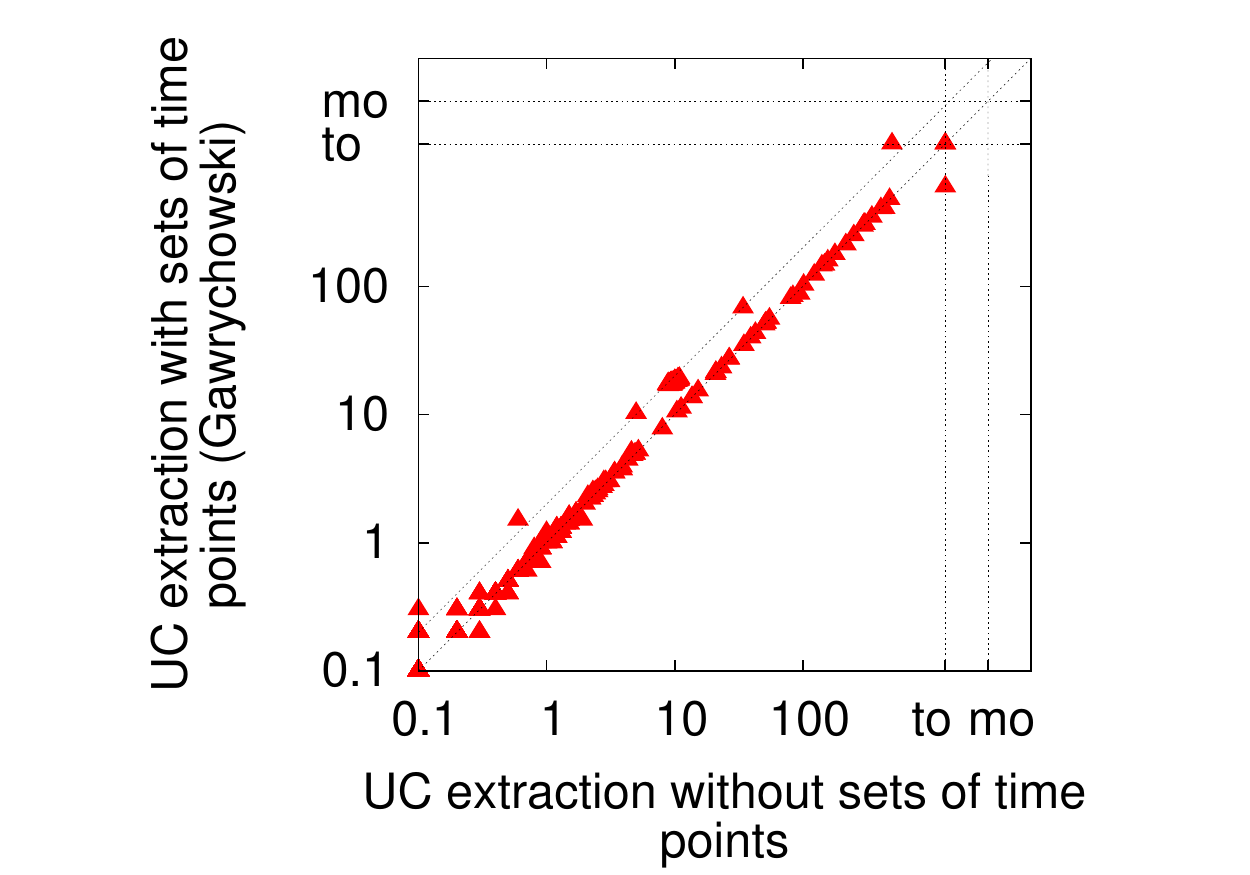} &
\tabfield{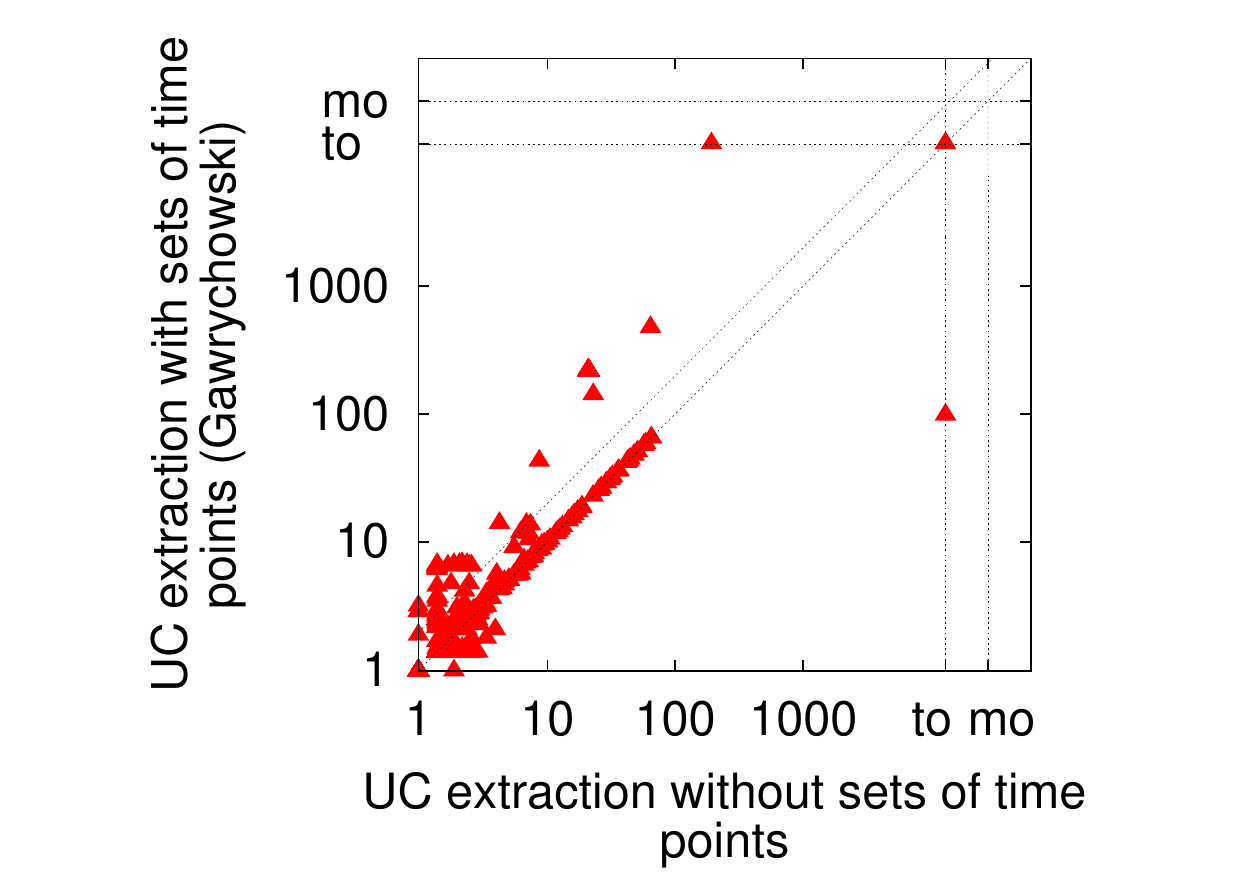} &
\tabfield{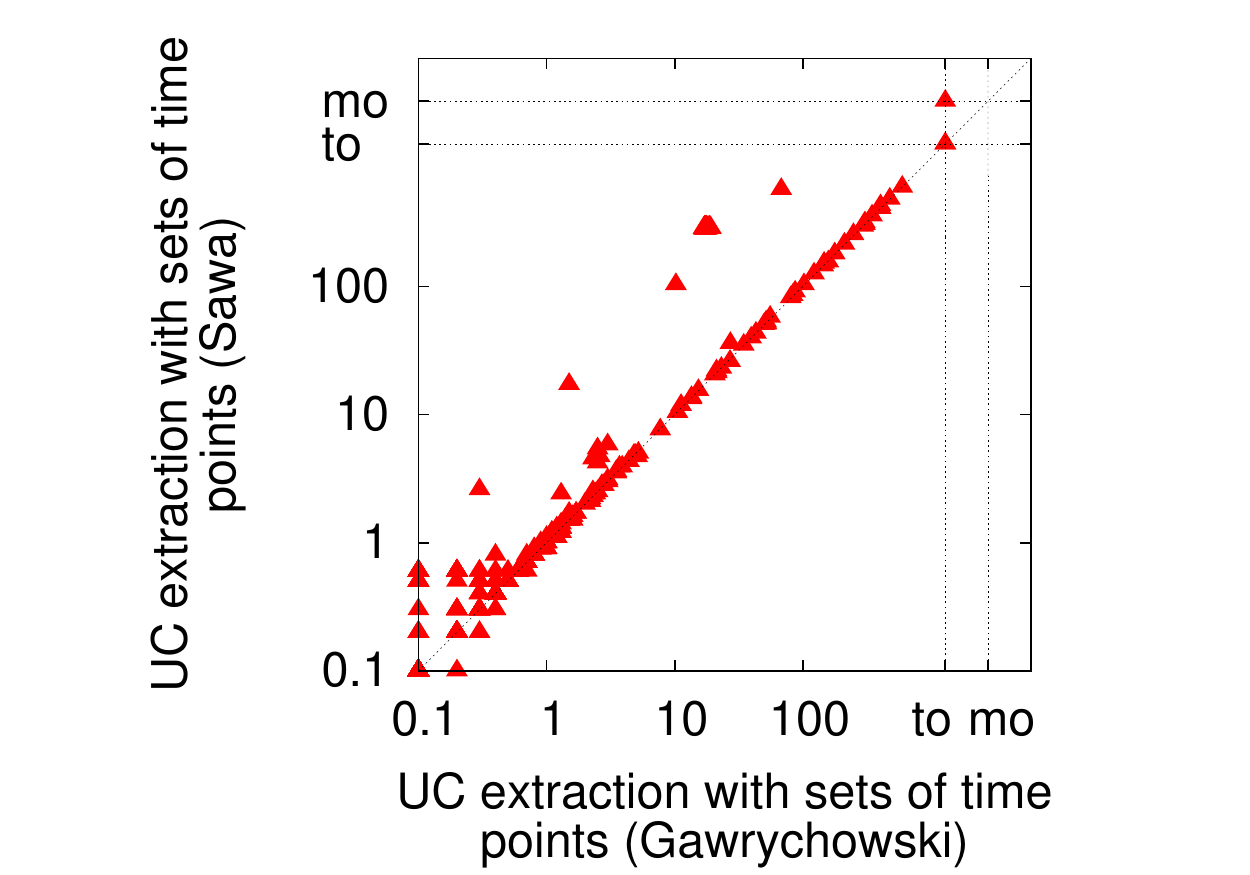} &
\tabfield{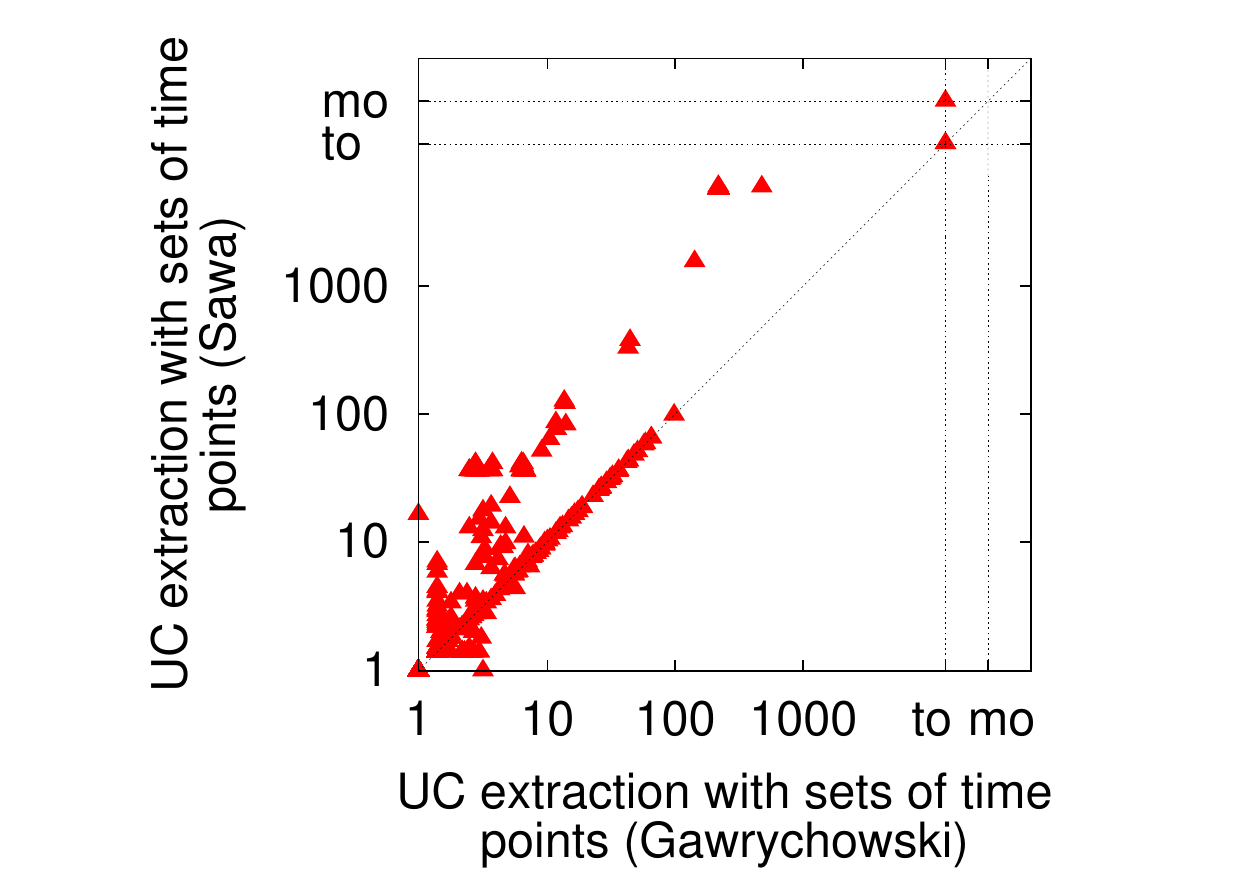} \\
\\[-0.5ex]
(a) run time [sec] & (b) memory [MB] & (c) run time [sec] & (d) memory [MB]
\\[-0.5ex]
\end{tabular}
\caption{\label{fig:overheadall}Overhead of \uc extraction with sets of time points: (a) and (b) show run time and memory for \uc extraction with sets of time points using Gawrychowski's algorithm (y-axis) versus \uc extraction without sets of time points (x-axis). (c) and (d) compare run time and memory for Sawa's algorithm (y-axis) and Gawrychowski's algorithm (x-axis) for \uc extraction with sets of time points. The off-center diagonal in (a) and (b) shows where $y = 2 x$.}
\end{figure}
}

{
\newcommand{\tabcolumnheader}[1]{%
\begin{minipage}{0.02\linewidth}%
\begin{sideways}%
{#1}%
\end{sideways}%
\end{minipage}%
}

\newcommand{\tabcolumnheadertwo}[1]{%
\begin{minipage}{0.035\linewidth}%
\begin{sideways}%
\begin{minipage}{5.5em}%
\begin{center}
{#1}%
\end{center}
\end{minipage}%
\end{sideways}%
\end{minipage}%
}

\newcommand{\tabcolumnheaderthrees}[1]{%
\begin{minipage}{0.0475\linewidth}%
\begin{sideways}%
\begin{minipage}{5.5em}%
\begin{center}
{#1}%
\end{center}
\end{minipage}%
\end{sideways}%
\end{minipage}%
}

\newcommand{\tabcolumnheaderthreew}[1]{%
\begin{minipage}{0.05\linewidth}%
\begin{sideways}%
\begin{minipage}{5.5em}%
\begin{center}
{#1}%
\end{center}
\end{minipage}%
\end{sideways}%
\end{minipage}%
}

\begin{table}[t]
\centering
{\tiny
\begin{tabular}{||@{\hspace{0.25em}}l@{\hspace{0.25em}}|@{\hspace{0.25em}}l@{\hspace{0.25em}}||@{\hspace{0.2em}}*{31}{c@{\hspace{-0.4em}}|@{\hspace{0.2em}}}c@{\hspace{-0.2em}}||}
\hline
\hline
category &
family &
\tabcolumnheader{$\zeroset$} &
\tabcolumnheader{$\{0,1\}$} &
\tabcolumnheader{$\{0,1,2\}$} &
\tabcolumnheader{$\{0,2\}$} &
\tabcolumnheader{$\{1\}$} &
\tabcolumnheader{$\{1,2\}$} &
\tabcolumnheader{$\{1,2,3\}$} &
\tabcolumnheader{$\{1,2,3,4\}$} &
\tabcolumnheader{$\{1,3\}$} &
\tabcolumnheader{$\{1,4\}$} &
\tabcolumnheader{$\{2\}$} &
\tabcolumnheader{$\{2,3\}$} &
\tabcolumnheader{$\{2,3,4\}$} &
\tabcolumnheader{$\{3\}$} &
\tabcolumnheader{$\{3,4\}$} &
\tabcolumnheader{$\{4\}$} &
\tabcolumnheader{$\allnats$} &
\tabcolumnheader{$\allnats + 1$} &
\tabcolumnheader{$\allnats + 2$} &
\tabcolumnheader{$\allnats + 3$} &
\tabcolumnheader{$\allnats + 4$} &
\tabcolumnheader{$\allnats + 5$} &
\tabcolumnheader{$\allnats + 6$} &
\tabcolumnheaderthrees{$\allnats + 7,$\\[-0.5ex]$\ldots,$\\$\allnats + 10$} &
\tabcolumnheaderthrees{$\fslsone{4}{0},$\\[-0.5ex]$\ldots,$\\$\fslsone{4}{5}$} &
\tabcolumnheadertwo{$\{\fls{4}{1},$ $\fls{4}{2}\}$} &
\tabcolumnheaderthreew{$\{\fls{4}{1},$ $\fls{4}{2},$ $\fls{4}{3}\}$} &
\tabcolumnheadertwo{$\{\fls{4}{2},$ $\fls{4}{3}\}$} &
\tabcolumnheaderthreew{$\{\fls{4}{2},$ $\fls{4}{3},$ $\fls{4}{4}\}$} &
\tabcolumnheadertwo{$\{\fls{4}{3},$ $\fls{4}{4}\}$} &
\tabcolumnheaderthrees{$\fslsone{5}{0},$\\[-0.5ex]$\ldots,$\\$\fslsone{5}{5}$} &
\tabcolumnheaderthrees{$\fslsone{12}{0},$\\[-0.5ex]$\ldots,$\\$\fslsone{12}{12}$}
\\
\hline
\hline
&
\benchmark{alaska\_lift} &
\yes&
\yes&
&
&
\yes&
\yes&
&
&
&
&
\yes&
&
&
&
&
&
\yes&
\yes&
&
&
&
&
&
&
&
&
&
&
&
&
&
\\
\cline{2-34}
\benchmark{application} &
\benchmark{anzu\_genbuf} &
\yes&
&
&
&
\yes&
&
&
&
&
&
&
&
&
&
&
&
\yes&
\yes&
\yes&
\yes&
\yes&
\yes&
\yes&
&
\yes&
\yes&
\yes&
\yes&
\yes&
\yes&
&
\\
\cline{2-34}
&
\benchmark{forobots} &
\yes&
\yes&
&
&
\yes&
&
&
&
&
&
&
&
&
&
&
&
\yes&
\yes&
\yes&
\yes&
\yes&
\yes&
&
&
&
&
&
&
&
&
&
\\
\hline
\hline
&
\benchmark{schuppan\_O1formula} &
\yes&
&
&
&
\yes&
&
&
&
&
&
&
&
&
&
&
&
&
&
&
&
&
&
&
&
&
&
&
&
&
&
&
\\
\cline{2-34}
\benchmark{crafted} &
\benchmark{schuppan\_O2formula} &
\yes&
&
&
&
&
&
&
&
&
&
&
&
&
&
&
&
\yes&
\yes&
&
&
&
&
&
&
&
&
&
&
&
&
&
\\
\cline{2-34}
&
\benchmark{schuppan\_phltl} &
\yes&
&
&
&
&
&
&
&
&
&
&
&
&
&
&
&
\yes&
\yes&
\yes&
\yes&
\yes&
\yes&
\yes&
\yes&
&
&
&
&
&
&
&
\\
\hline
\hline
&
\benchmark{rozier\_formulas} &
\yes&
\yes&
\yes&
\yes&
\yes&
\yes&
\yes&
\yes&
\yes&
\yes&
\yes&
\yes&
\yes&
\yes&
\yes&
\yes&
\yes&
\yes&
\yes&
\yes&
\yes&
\yes&
&
&
&
&
&
&
&
&
&
\\
\cline{2-34}
\raisebox{1.5ex}[-1.5ex]{\benchmark{random}} &
\benchmark{trp} &
\yes&
&
&
&
\yes&
&
&
&
&
&
&
&
&
&
&
&
\yes&
\yes&
\yes&
\yes&
&
&
&
&
&
&
&
&
&
&
\yes&
\yes\\
\hline
\hline
\end{tabular}
}
\caption{\label{tab:slsdistribution}Occurrences of sets of time points in \ucs: A \yes in a field indicates that a subformula in a \uc of that benchmark family is assigned that set of time points.}
\end{table}
}

\ifqaplsubmission
\else
\paragraph{Discussion}
\fi

Our data show that extraction of \ucs with sets of time points is possible with quite acceptable overhead in run time and memory usage (Fig.~\ref{fig:overheadall} (a), (b)). In particular, out of the 698 instances we considered with \uc extraction without sets of time points, a \uc was obtained for 611. With sets of time points enabled one instance more\footnote{For this instance the run time with sets of time points is just below the time limit.} and one instance less are solved. An analysis by category (for plots see \refmoreplots) shows that the run time (resp., memory) overhead for almost all instances of the \benchmark{application} category is at most 50 \% (resp., 100 \%) for \uc extraction with sets of time points using Gawrychowski's algorithm over \uc extraction without sets of time points.

Sets of time points often provide helpful information. For some subfamilies of the \benchmark{anzu\_genbuf} and \benchmark{trp} families they show that some subformulas are required only every 4th, 5th, or 12th time point. For an instance of the \benchmark{forobots} family they make it clear that only the first two time points are relevant, \ifqaplsubmission although \else even though \fi some of the subformulas involved are $\fgloballyname$ subformulas. For the \benchmark{schuppan\_phltl} family (a temporal version of the pigeon hole problem\ifqaplsubmission\else\xspace(e.g., \cite{DBLP:series/faia/2009-185})\fi; $n$ pigeon holes are turned into a single pigeon hole over $n$ time points) they indicate how the conditions of mutual exclusivity for the hole are invoked one after the other.

Gawrychowski's algorithm \cite{PGawrychowski-CIAA-2011} has better worst case complexity than Sawa's algorithm \cite{ZSawa-FundamentaInformaticae-2013}. We also found it easier to understand and implement. On our benchmarks Gawrychowski's algorithm tends to perform better than Sawa's algorithm (Fig.~\ref{fig:overheadall} (c) and (d)), especially when the NFAs become larger.

\section{Conclusions}
\label{sec:conclusions}

In this paper we showed how to obtain information on the time points at which subformulas of a \uc for LTL are required for unsatisfiability, providing useful information in many cases and leading to a more fine-grained notion of \uc than in \cite{VSchuppan-SCP-2012}.
We demonstrated with an implementation in \trp that \ucs with sets of time points can be extracted efficiently.
Potential future work includes extending the computation of sets of time points to tableau-based \uc extraction for LTL such as \cite{FHantryMHacid-FLACOS-2011} and exploring whether computation of sets of time points is feasible for BDD-based algorithms via, e.g., \ifqaplsubmission\cite{TJussilaCSinzABiere-SAT-2006}\else\cite{CSinzABiere-CSR-2006,TJussilaCSinzABiere-SAT-2006}\fi.
Other questions are how to apply the idea of sets of time points to unrealizable cores for LTL (e.g., \cite{VSchuppan-SCP-2012}) or to branching time temporal logics.
One could also investigate obtaining sets of time points by solving a system of constraints over sets of time points based on Lemmas \ref{thm:initialclauseslabeledzeroset}, \ref{thm:targetclauselabeledsubsetofsourceclause} rather than the approach based on Parikh images explored here.
Finally, it would be interesting to see whether/how minimal or minimum sets of time points can be obtained, where $\le$ is set inclusion (rather than syntactic expression size).

{\small
\paragraph{Acknowledgements}
\addcontentsline{toc}{section}{Acknowledgements}

I thank B.~Konev and M.~Ludwig for making \trp and \tspass including their LTL translators available. I also thank A.~Cimatti for bringing up the subject of temporal resolution and for pointing out that the resolution graph can be seen as a regular language acceptor. Initial parts of the work were performed while working under a grant by the Provincia Autonoma di Trento (project EMTELOS).
}


\ifqaplsubmission
\bibliographystyle{eptcs}
\bibliography{trpuc-paper-short}
\else
\begin{sloppypar}
\printbibliography[heading=bibintoc]
\end{sloppypar}
\fi
\ifnoappendix
\else
\clearpage
\appendix

\input{formal-ltli.tex}
\input{formal-coreextractioni.tex}
\input{moreexamples.tex}
\input{moreplots.tex}
\fi
\end{document}